\documentclass[fleqn,usenatbib]{mnras}

\usepackage{newtxtext,newtxmath}
\usepackage[T1]{fontenc}
\usepackage{ae,aecompl}

\usepackage{ulem}

\usepackage{graphicx}   % Including figure files
\usepackage{amsmath}    % Advanced maths commands
\usepackage{cite}

\usepackage{multirow}
\usepackage{booktabs}
\usepackage{tabularx}

\newpage

\graphicspath{{./figures/}}

\title[Photoionisation in spiral arm clouds]{Photoionising feedback in spiral arm molecular clouds}

\author[T. J. R. Bending, C. L. Dobbs and M. R. Bate]{
Thomas J. R. Bending,
Clare L. Dobbs,
Matthew R. Bate
\\
School of Physics and Astronomy, University of Exeter, Stocker Road, Exeter EX4 4QL, UK\\
}

\date{Accepted XXX. Received YYY; in original form ZZZ}

\pubyear{2020}

\begin{document}
\label{firstpage}
\pagerange{\pageref{firstpage}--\pageref{lastpage}}
\maketitle

\begin{abstract}
    We present simulations of a 500 pc$^2$ region, containing gas of mass 4 $\times$ 10\textsuperscript{6} M$_\odot$, extracted from an entire spiral galaxy simulation, scaled up in resolution, including photoionising feedback from stars of mass \textgreater 18 M$_\odot$. Our region is evolved for 10 Myr and shows clustered star formation along the arm generating $\approx$ 5000 cluster sink particles $\approx$ 5\% of which contain at least one of the $\approx$ 4000 stars of mass \textgreater 18 M$_\odot$. Photoionisation has a noticeable effect on the gas in the region, producing ionised cavities and leading to dense features at the edge of the HII regions. Compared to the no-feedback case, photoionisation produces a larger total mass of clouds and clumps, with around twice as many such objects, which are individually smaller and more broken up. After this we see a rapid decrease in the total mass in clouds and the number of clouds. Unlike studies of isolated clouds, our simulations follow the long range effects of ionisation, with some already-dense gas, becoming compressed from multiple sides by neighbouring HII regions. This causes star formation that is both accelerated and partially displaced throughout the spiral arm with up to 30\% of our cluster sink particle mass forming at distances \textgreater 5 pc from sites of sink formation in the absence of feedback. At later times, the star formation rate decreases to below that of the no-feedback case.
    
\end{abstract}


%%%%%%%%%%%%%%%%%%%%%%%%%%%%%%%%%%%%%%%%%%%%%%%%%%

%%%%%%%%%%%%%%%%% BODY OF PAPER %%%%%%%%%%%%%%%%%%

\section{Introduction}
\label{se:intro}

The majority of stars form in giant molecular clouds (GMCs) and in turn many GMCs are found in complexes typically in galactic spiral arms.  Observations indicate that star formation rates are lower than simple theoretical predictions \citep{Zuckerman1974,Krumholz2007a,Evans2009,Semenov2017}, but how star formation is regulated is still an open question. The relative importance of processes external to GMCs (such as galactic potentials and external radiation fields) and internal processes (such as stellar feedback and turbulence) in star forming regions in particular is poorly understood. In particular different feedback effects are often explored on smaller scales, but in the absence of large scale processes such as spiral arms. In this work we focus  on one important feedback process, photoionisation, and simulate photoionising feedback in detail while also considering the galactic environment.

Observationally photoionising feedback and the resulting HII regions are thought to play a vital role in determining the sites of star formation through triggering. Most notably this is expected to happen via the collect and collapse model \citep{Elmegreen1977}, where shells of gas are swept up in front of ionisation fronts (IFs) and fragment into star forming clumps, which has been observed by \citet{Deharveng2005}, \citet{Pomares2009} and \citet{Zavagno2010}. Within HII regions bright rimmed clouds are often created by radiation-driven implosion \citep{Bertoldi1989,Morgan2008}. Both of these mechanisms that can potentially trigger star formation have been observed on scales of up to several pc, yet surveys of HII regions show that OB associations are capable of creating shells many tens of pc across \citep{Anderson2014,Hindson2016}. This suggests that there could be further-reaching, but harder-to-observe, effects of photoionisation.

From numerical simulations, there have also been many studies of the role of stellar feedback in individual molecular clouds. These have shed light on the role and relative importance of different feedback mechanisms in different environments. However, there remain many questions relating to their collective role, especially over larger size and longer time scales. Single cloud studies have the advantage that the gas and feedback are well resolved, but have the disadvantage that once the feedback reaches the edge of the cloud it is completely free to escape outside the computational domain. For many of these simulations, there is one single massive star, for example \citet{Ali2018,Ali2019} include a 34 M$_{\odot}$ star, \citet{Haid2018} examine the effects from three different mass stars, \citet{Geen2016} investigate clouds with different ionising sources. Others such as \citet{Dale2014} and \citet{Zamora-Aviles2019} consider more massive ($10^5$ M$_{\odot}$) clouds including multiple sources. 

These simulations tend to find that photoionisation has a major role in disrupting the cloud, even if the star formation rate (SFR) is still high compared to those observed \citep{Dale2014}. As would be expected, smaller clouds subject to higher ionising sources are more efficently dispersed by ionisation \citep[e.g.][]{Gavagnin2017}. \citet{Dale2014} investigate the combined effect of both ionisation and winds, concluding that ionisation is dominant over winds, however, winds can still have a significant role in shaping the morphology of low mass clouds. \citet{Haid2018} also conclude that ionisation tends to be more dominant, although winds can be more effective when interacting with the warm neutral medium. \citet{Kim2018} suggest that radiation pressure may be more important for denser and more massive clouds. Several studies suggest that feedback from ionisation and winds is able to disrupt clouds before supernovae take effect \citep{Rogers2013,Walch2015,Rey-Raposo2017,Kim2019}, but again these tend to be studies of isolated clouds where, once ionisation or winds have dispersed the cloud, the supernovae bubbles are not contained within the cloud so have little effect. Although these works demonstrate that on the scale of a single cloud photoionisation has the ability destroy clouds and shut down star formation, a key  question of what happens to the ejected gas remains.

On much larger scales, whole galaxy simulations have the capacity to study feedback in clouds situated within a galaxy, and include galaxy scale phenomena such as spiral arms, but lack the resolution to study detailed feedback. There is a large body of work which includes supernova feedback, which can be modelled adequately if the resolution is such that energy can be inserted into a resolvable number of particles or grid cells \citep{DallaVecchia2008,DallaVecchia2012,Dobbs2011,Hopkins2018a} and in these simulations the effects of overcooling, which can reduce the effectiveness of feedback, are generally overcome or mitigated. Most simulations show that feedback, either modelled as supernovae, or collectively from multiple forms, is able to disperse clouds in a galaxy \citep{Dobbs2011}, induce a velocity dispersion \citep{Joung2009,Seifried2018}, eject gas outside the plane of the disc \citep{DeAvillez2005,Kim2019,Hill2012,Gatto2017} and lead to a more realistic population of clouds and star formation rates \citep{Dobbs2011,Grisdale2018,Kim2013,Hopkins2011}. However, these simulations miss the detailed evolution of feedback, and for example, the expected impact on the clouds of the different types of feedback suggested by isolated cloud simulations \citep{Dale2014,Geen2016,Ali2019,Zamora-Aviles2019}.

Other forms of feedback, such as ionisation, are often considerably simplified if included in whole galaxy simulations. Typically ionisation is included by computing the Str{\"o}mgren radius surrounding a sink particle, and then the gas within this radius set to 10$^4$ K \citep{Hopkins2014,Renaud2013,Baba2017}. Simulations of dwarf galaxies allow for higher resolution due to their lower mass and \citet{Emerick2018} show that photoionisation in addition to supernovae can reduce the star formation rate in dwarf galaxies by up to a factor of five. However, in isolated GMC simulations we typically see a far more modest change. 

An alternative approach is to perform simulations on an intermediate scale, which allows the simulation domain to include multiple clouds, as well as a rigorous model for pre-supernovae feedback. \citet{Walch2015,Girichidis2016,Gatto2017} use a 500 pc\textsuperscript{2} box generated using a Gaussian profile for gas density in the z direction and a uniform density in the x-y plane. In addition these simulations include ISM chemistry and an external gravitational field. They model winds and supernovae and find that, similar to the single cloud models, the winds appear to have a significant impact on the clouds before supernova occur. \citet{Butler2017} model kpc sized regions of galaxies investigating the effects of photodissociation, ionisation and supernovae, finding that supernovae and ionisation have more modest effects compared to photodissociation. Although they don't investigate star formation, \citet{Vandenbroucke2019} model ionisation in vertical sections of the disc and show that ionisation is capable of driving turbulence in the gas.

In this paper we re-simulate regions from a galaxy disc simulation and investigate the impact of photoionisation on molecular clouds and star formation. We maintain the galactic physics from the parent simulation: heating, cooling, and galactic potential. Additionally we model photoionisation with a ray-tracing type method which takes into account gas distribution between particles in the simulation and nearby ionising sources. Our simulated regions are large enough that we can follow the propagation of ionisation beyond individual molecular clouds, through low density diffuse ISM, and see the impact of ionisation fronts as they expand to reach the neighbouring cold atomic or molecular gas clouds. In Section \ref{sec:methods} we present our numerical methods including our method for sampling massive stars, and implementation of ionisation. In Section \ref{sec:results} we present our results, including the effect of ionisation on the overall morphology of the gas, the star formation rates in different models and the properties of the resultant clouds and stellar clusters.

\section{Numerical Methods}
\label{sec:methods}

\begin{table}
    \centering
    \caption{Summary of initial conditions details}
    \label{tab:initital_conditions}
    %\begin{tabular}{|p{1.25cm}|p{1.0cm}|p{1.4cm}|p{1.2cm}|p{1.2cm}|} % five columns, alignment for each
    \begin{tabularx}{\columnwidth}{p{0.9cm}p{1.1cm}p{1.4cm}p{1.5cm}p{1.4cm}}
        \toprule
        IC ref & Mass (M$_\odot$) & x-y plane \newline size (kpc) & Particle Mass (M$_\odot$) & Number of Particles\\
        \toprule
        SR & $4\times10^6$ & 0.5 $\times$ 0.5 & 1.00 & 4 $\times10^6$\\
        HR & $4\times10^6$ & 0.5 $\times$ 0.5 & 0.37 & 10 $\times10^6$\\
        IA & $5\times10^6$ & 1.0 $\times$ 1.0 & 1.00 & 5 $\times10^6$\\
        \bottomrule
    \end{tabularx}
\end{table}

The calculations presented in this paper were performed using the three-dimensional smooth particle hydrodynamics (SPH) code, {\sc sphNG}.  The code originated from W. Benz \citep{Benz1990sph-review,Benz1990}, but has since been substantially modified  \citep*{Bate1995,Price2007b} and has been parallelised using both {\sc OpenMP} and the message passing interface (MPI). The code also includes implementations of radiative transfer, non-ideal MHD, and dust drag, but these elements are not relevant for this paper.

\subsection{Extracting initial conditions}
\label{sec:ICs}

We extract square regions from galaxy scale simulations by \citet{Dobbs2013}. The simulations of \citet{Dobbs2013} followed molecular cloud evolution in a spiral galaxy, by modelling a gaseous disc with a galactic potential. The simulations included ISM heating and cooling, self gravity and stellar feedback modelled as supernovae feedback. We extract initial conditions from a time of 240 Myr, whereby the simulations have reached a rough equilibrium in terms of molecular cloud properties, and the amounts of gas in different phases \citep{Dobbs2011}. We use regions of 0.5-1 kpc in size at full depth through the x-y plane, which is the maximum size for which we are able to resolve the ionisation, given our particle numbers and according to our resolution tests (Section \ref{sec:resolution} and Appendix \ref{app:mass_resolution}). Whilst \citet{Dobbs2015} trace SPH particles back in time to select their initial conditions, we do not do this here since we only run these simulations for relatively short timescales and the shape of the region does not change substantially.

We increase the resolution of the extracted initial conditions using a method similar to \citet{Rey-Raposo2014}. \citet{Rey-Raposo2014} distribute $N$-1 new particles within $2h$ of each original particle, where $h$ is the SPH smoothing length and reduce the particle mass by a factor of $N$. The radial positions of their added particles follow the SPH smoothing kernel's density distribution. 

We instead use a spherical grid created by placing $N$-1 particles around the original particle. The grid has a radius of twice the smoothing length (i.e. the radius of compact support of the kernel). The separation of particles in this grid decreases with radius following the normalised inverse cube of the SPH kernel function; this ensures that density falls off with radius in line with the kernel. The grid is made up of a set of concentric shells and is described in detail in Appendix \ref{app:part_resolution}.

\begin{figure}
  \begin{center}
    \centering
    \includegraphics[width=0.9\columnwidth]{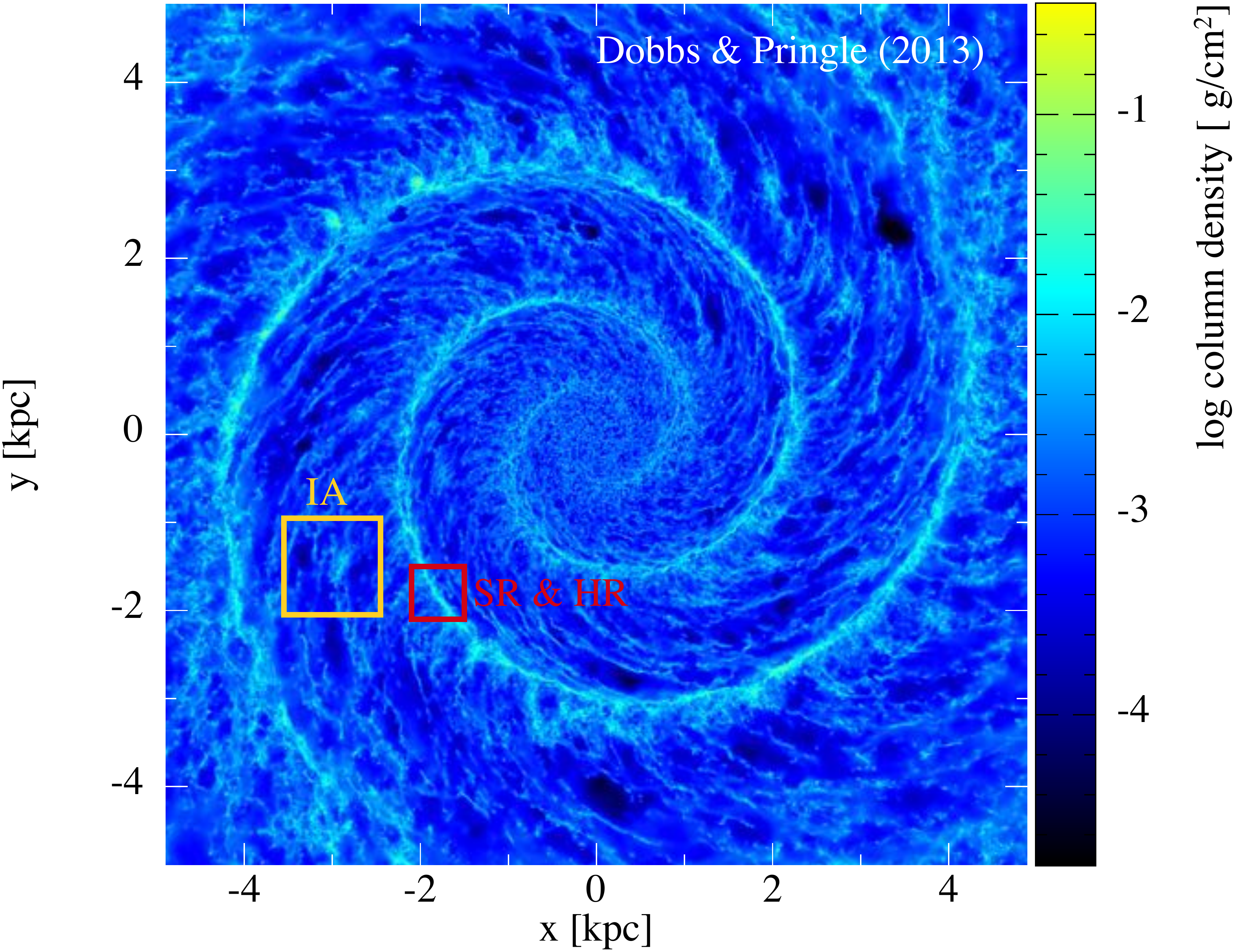}
  \end{center}
  \begin{center}
    \includegraphics[width=0.9\columnwidth]{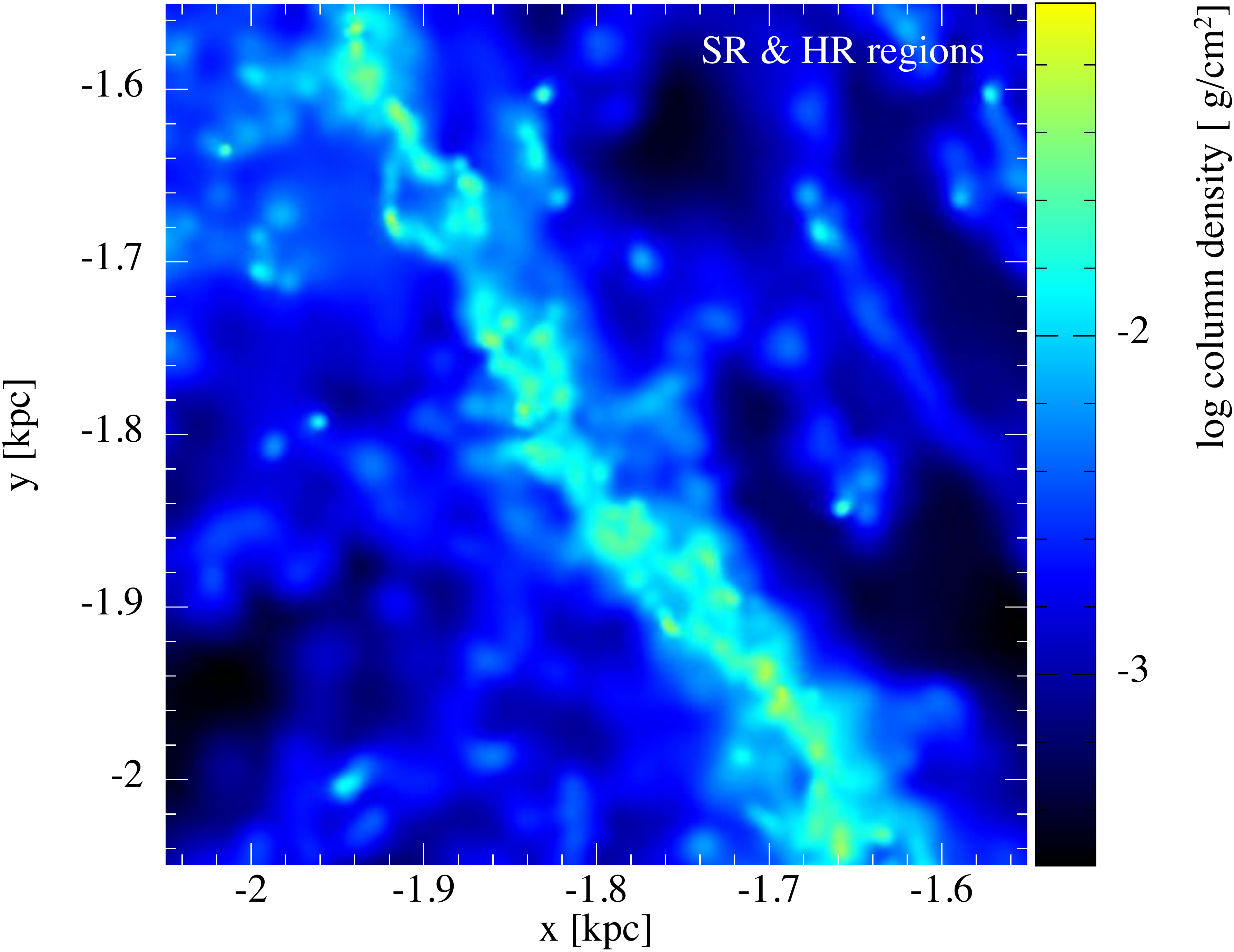} 
  \end{center}
  \begin{center}
    \includegraphics[width=0.9\columnwidth]{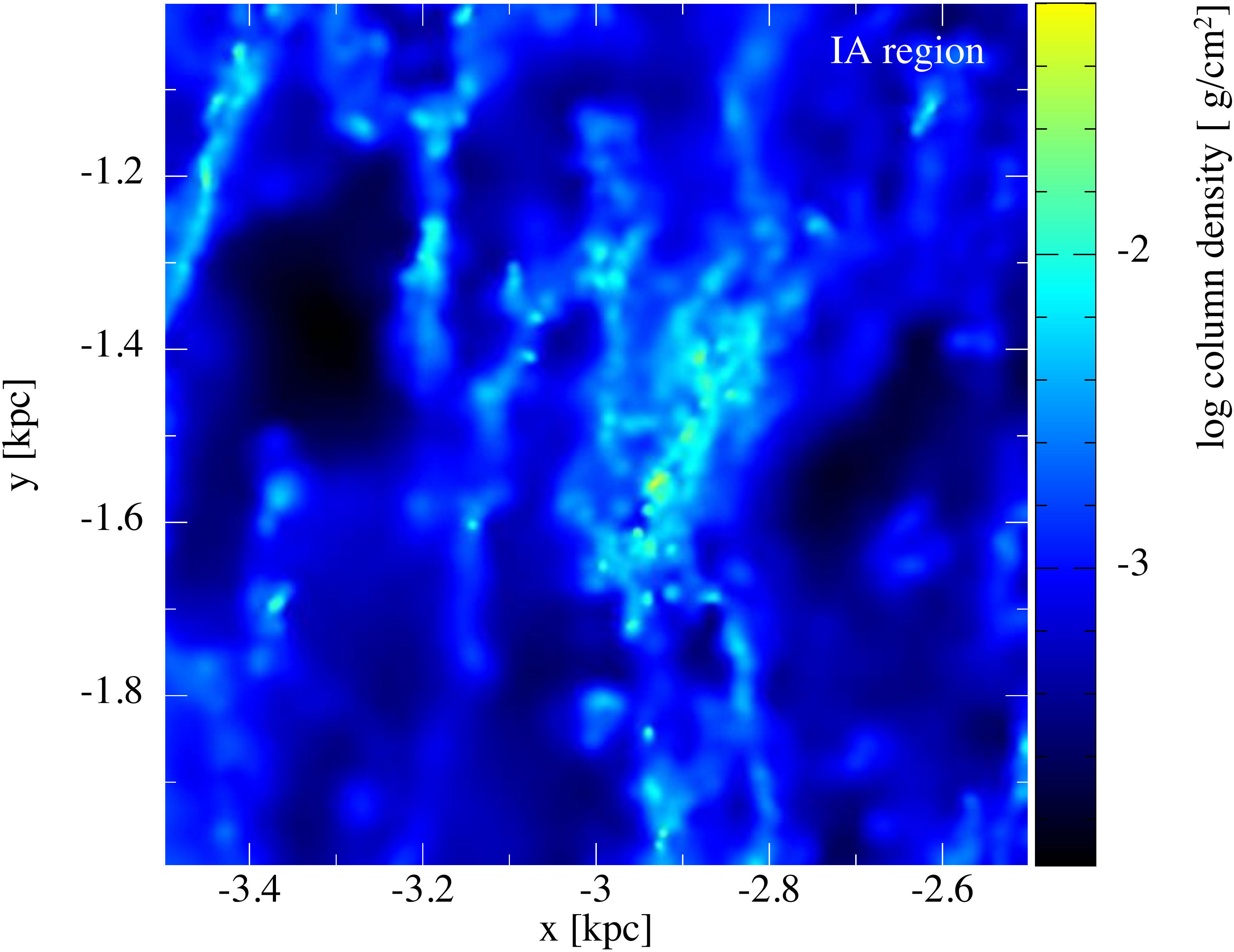}
  \end{center}
  \caption{The top panel shows a galaxy simulation from \citet{Dobbs2013} that has evolved for 240 Myr. The smaller red box is the spiral arm region enlarged in the middle panel and the larger yellow box is the inter-arm region enlarged in the bottom panel. The resolution of these boxes is increased as described in Section \ref{sec:ICs}.}
  \label{fig:ICs}
\end{figure}

All of our simulations use one of three sets of initial conditions extracted in this way: a region with a spiral arm passing through its centre (SR), the same region at a higher resolution (HR) and a larger inter-arm region (IA). These are summarised in Table \ref{tab:initital_conditions}. The SR and IA initial conditions use N = 311 and the HR initial conditions use N = 823. The mass of each SPH particle in these simulations is then 1.00 M$_\odot$ and 0.37 M$_\odot$ respectively. An example of the extraction of the SR region is shown in Figure \ref{fig:ICs}.

\begin{table}
  \centering
  \caption{Summary of the details of the simulations. Column 3 gives the accretion radius of the sink particles.}
  \label{tab:simulations}
  \begin{tabularx}{\columnwidth}{p{1.11cm}p{1.05cm}p{0.77cm}p{4.2cm}} 
    \toprule
    Run & Feedback & Radius (pc) & Feature varied \\
    \toprule
    SR        & none     & 0.78 & \multirow{2}{4cm}{Fiducial runs} \\
    SR\_ion   & ionising & 0.78 &  \\
    \addlinespace[0.2cm]
    SR\_50\%  & ionising & 0.78 & Star formation efficiency 50\% \\
    SR\_los   & ionising & 0.78 & No distance limit for photons \\
    SR\_alt   & ionising & 0.78 & Alternative star sampling \\
    \addlinespace[0.2cm]
    HR        & none     & 0.78 & \multirow{2}{4cm}{High resolution}  \\
    HR\_ion     & ionising &  & \\
    \addlinespace[0.2cm]
    IA        & none     & 0.78 & \multirow{2}{4cm}{inter-arm region}  \\
    IA\_ion     & ionising & 0.78 & \\
    \addlinespace[0.2cm]
    SR\_2    & none     & 0.45 & \multirow{2}{4cm}{Medium accretion radius}  \\
    SR\_2\_ion  & ionising & 0.45 & \\
    \bottomrule
  \end{tabularx}
\end{table}

\subsection{Sink particles and sampling of massive stars}
\label{se:sinksandsampling}

\begin{table}
  \centering
  \caption{Data from \citet{Sternberg2003} along with the bin boundaries used. The upper bin boundaries correspond to the lower bin boundaries for the previous bin number. $Q_h$ is the representative ionising flux of a star in a given bin, we ignore ionisation from stars in the 16\textsuperscript{th} bin.}
  \label{tab:Stern}
  \setlength\tabcolsep{3.5pt} % default value: 6pt
  \begin{tabularx}{\columnwidth}{c c c c c}
    \toprule
    Bin    & Spectral   & Representative          & Upper\:bin           & log\:$Q_h$  \\
    number &  type      & bin mass [M$_\odot$]    & boundary [M$_\odot$] & [log s$^{-1}$] \\
    \toprule
    1      & O3\:\:\:   & 87.6                    & 107.5              & 49.87                \\ 
    2      & O4\:\:\:   & 68.9                    & 72.8               & 49.68                \\ 
    3      & O4.5       & 62.3                    & 65.5               & 49.59                \\ 
    4      & O5\:\:\:   & 56.6                    & 59.5               & 49.49                \\ 
    5      & O5.5       & 50.4                    & 53.7               & 49.39                \\ 
    6      & O6\:\:\:   & 45.2                    & 47.7               & 49.29                \\ 
    7      & O6.5       & 41.0                    & 43.0               & 49.18                \\ 
    8      & O7\:\:\:   & 37.7                    & 39.3               & 49.06                \\ 
    9      & O7.5       & 34.1                    & 36.0               & 48.92                \\ 
    10\:\: & O8\:\:\:   & 30.8                    & 32.4               & 48.75                \\ 
    11\:\: & O8.5       & 28.0                    & 29.4               & 48.61                \\ 
    12\:\: & O9\:\:\:   & 25.4                    & 26.7               & 48.47                \\ 
    13\:\: & O9.5       & 23.3                    & 24.3               & 48.26                \\ 
    14\:\: & B0\:\:\:   & 21.2                    & 22.2               & 48.02                \\ 
    15\:\: & B0.5       & 19.3                    & 20.3               & 47.71                \\ 
    16\:\: & All other  & \:\:0.3                 & 18.4               & N/A            \\ 
    \bottomrule
    \end{tabularx}
\end{table}

On this scale each sink particle represents a cluster or sub-cluster of stars, hereafter when we refer to sink particles we are referring to cluster sink particles. The sink particles form as described in \citet{Bate1995}, whereby gas particles are tested once they exceed a threshold density, and a sink particle is created if the following conditions are satisfied for the $\approx$ 50 SPH neighbours:

\begin{enumerate}
\item ratio of thermal to gravitational energies is $\leq \frac{1}{2}$.
\item the sum of the thermal and rotational energies over the gravitational energy is $\leq 1$.
\item total energy of the particles is negative.
\item divergence of the particles' accelerations must be negative.
\end{enumerate}

The critical number density, the maximum density at which the Jeans mass can be resolved \citep{Bate1997}, is 1.2 $\times\ 10^4$ cm$^{-3}$ at a temperature of 30 K in the HR runs. We use this as the threshold density in all our simulations, however, we find that gas routinely exceeds this threshold without forming sink particles, as it does not meet the other criteria listed above. We therefore also introduce a second criteria of 1.2 $\times\ 10^6$ cm$^{-3}$ where automatic sink creation occurs, preventing gas densities becoming exceptionally high. In reality we would expect that star formation would occur at these densities, but potentially we lack the resolution such that gas in a region matches all the criteria listed above.

The Jeans radius for gas at this density in the SR initial conditions is 0.45 pc which is the accretion radius we used in the SR\_2 runs. A larger accretion radius, and hence sink particle resolution, of 0.78 pc is used in the SR and HR runs.

The close approach of sink particles leads to very short timesteps and hence considerable computational expense, particularly in runs containing many sink particles. To avoid this we use a relatively large separation criterion at which sink particles are merged of 0.01 pc. Since stellar dynamics are of secondary interest in this work, we consider the resulting decrease in computational cost a reasonable payoff for the loss in sink dynamics.

To account for ionisation we need to track the formation of massive stars in our simulation. To do this, we sample 3$\times10^6$ M$_\odot$ using a \citet{Kroupa2001} initial mass function (IMF) into mass bins using the approach presented by \citet{Sormani2017}. Our bin boundaries are defined such that the representative mass of each bin corresponds to the 15 luminosity class V stars presented by \citet{Sternberg2003}, with a 16\textsuperscript{th} bin for all less massive stars. Ignoring the 16\textsuperscript{th} bin gives us a pre-sampled list of bin numbers (1-15) of stars above 18 M$_\odot$, where each bin has an associated ionising flux. We have re-created the relevant parts of the table from \citet{Sternberg2003} in Table \ref{tab:Stern} along with our bin boundaries.

Similarly to \citet{Geen2018} we define $\Delta M_{i}$, the massive star injection interval, as the total mass of all stars in the pre-sampled list (stars \textless 18 M$_\odot$) over the total mass of the sample (3$\times10^6$ M$_\odot$). Every time the total sink particle mass in a simulation passes a multiple of $\Delta M_{i}$, we inject the next massive star in the pre-sampled list. Each injected massive star is assigned to the sink particle with the greatest mass made up of stars less than 18 M$_\odot$; if there is no sink particle massive enough to accommodate the star then injection is delayed. $\Delta M_{i}$ = 305 M$_\odot$ for all simulations except SR\_alt that uses an alternative sample of stars, of the same total mass, for which $\Delta M_{i}$ = 301 M$_\odot$.

\subsection{Heating, cooling and galactic potential}

We use the same galactic potential, heating and cooling as \citet{Dobbs2013}. The potential includes a logarithmic component which produces a flat rotation curve \citep{binney2008galactic} and a two armed spiral perturbation \citep{Cox2002}. Full details of the potentials are also provided in \citet{Dobbs2006}.

The heating and cooling of the ISM are described in \citet{Dobbs2008}, which is based on the work of \citet{Glover2007}. The method accounts for the following cooling processes: fine-structure emission lines of C\textsuperscript+,O and Si\textsuperscript+, ro-vibrational emission from H\textsubscript2, gas-grain energy transfer and recombination on grain surfaces. We also consider collisional dissociation of H$_2$, the collisional ionisation of atomic hydrogen and emission from atomic resonance lines and bremsstrahlung. The heating processes include, photoelectric emission from dust grains, H$_2$ photodissociation, and the pumping of excited vibrational states of H$_2$ by the background ultraviolet field or during the formation process of the molecules.

Self gravity is included, but, unlike \citet{Dobbs2013}, we do not include supernovae -- instead we model photoionising feedback. We use an assumed ionised gas temperature of $10^4$ K for gas ionised as a result of photoionisation. 

\subsection{Photoionisation}
\label{sec:ionmethod}

\subsubsection{Theory}

The ionisation of a uniform-cloud cloud of HI was studied first by \citet{Stromgren1939}, then by \citet{Kahn1954}, and subsequently by many others.  \citet{Kessel-Deynet2000} and \citet{Dale2007b} both presented methods for modelling photoionisation within the SPH method based on computing the Str{\"o}mgren radius along lines of sight.  \citet{Dale2007b} improved the basic method to take account of the fact that in a dynamical situation, neutral gas may enter an HII region from outside, and ionised gas may leave an HII region or be cut-off from the supply of photons that keep it from recombining.  The method presented below follows the basic method of \citet{Dale2007b}, with some modifications primarily to do with how the line of sight integrals are calculated.

If an ionising source is placed in a cloud of neutral hydrogen, the ionising photons cause an IF to propagate outward from the source at highly supersonic speed leaving behind an HII region.  This is known as the {\rm R-type} expansion phase.  Taking hydrogen with number density $n$, if the gas is fully ionised the number densities of ions and electrons are $n_{\rm i}=n_{\rm e}=n$.  The recombination rate per unit volume is then $\alpha n_{\rm i} n_{\rm e} = \alpha n^2$, where $\alpha$ is the recombination coefficient.  We consider any photon whose energy exceeds 13.6~eV as ionising and define the ionising photon luminosity as $Q_{\rm H}$ (measured in ionising photons emitted per unit time).  For a static cloud surrounding an ionising source with an arbitrary radial density profile and consisting of fully ionised hydrogen (i.e.\ within the HII region), the flux of ionising photons passing through radius $r$ can be written as
\begin{equation}
4 \pi r^2 F(r) = Q_{\rm H} - 4 \pi \int^{r}_{0} r^{\prime 2} n(r^\prime)^2 \alpha_{\rm B} {\rm d}r^\prime,
\label{eq:flux}
\end{equation}
which accounts for geometric dilution of the ionising photons and the photons required to balance recombination of ionised gas.  In setting the recombination rate, we make the `on-the-spot' (OTS) approximation which assumes that photons produced from recombinations directly to the hydrogen ground state are re-absorbed within the HII region.  The ionising photons produced by such recombinations are assumed to be absorbed elsewhere and not to contribute to the local ionisation equilibrium.  Thus, we take the temperature-dependent `case B' recombination coefficient to be $\alpha_{\rm B}=2.7\times 10^{-13}$~cm$^{3}$~s$^{-1}$.   This is justifiable when the optical depth of the HII region to secondary ionising photons is smaller than the dimensions of the HII region.  

As the IF expands into neutral gas, the integral in equation \ref{eq:flux} becomes equal to $Q_{\rm H}$ and the IF cannot proceed any further.  In a uniform-density cloud, $n=n_0$, the radius at which this happens is known as the Str\"omgren radius which is defined by integration of the above equation
\begin{equation}
R_{\rm S} = \sqrt[3]{\frac{3 Q_{\rm H}}{4 \pi n_0^2 \alpha_{\rm B}}}.
\label{eq:rs}
\end{equation}
Later expansion is driven by the pressure difference between the ionised gas and the surroundings and is known as D-type expansion.

Equation \ref{eq:flux} gives the net ionising flux as a function of radius if the intervening gas is fully ionised.  However, it will not capture the time dependence of the initial R-type expansion phase.  Neither will it treat the case of neutral gas entering the intervening region where some ionising photons will be used up ionising the intervening gas, or if the ionising source is reduced or blocked.

Instead, following \citet{Dale2007b},  we can write the number of ionising photons that passes through radius $r$ in time $\delta t$ in the form 
\begin{equation}
\begin{aligned}
\delta t \left( Q_{\rm H} - 4 \pi  \int^{r}_{0} r^{\prime 2}      \right. & \left.   n(r^\prime)^2  \alpha_{\rm B} {\rm d}r^\prime \right)  \\
 & - 4 \pi  \int^{r}_{0} r^{\prime 2} n(r^\prime) [1-H_{\rm II}(r^\prime)] {\rm d}r^\prime,
\label{eq:timedependent}
\end{aligned}
\end{equation}
where $H_{\rm II}$ is the ionisation fraction of the gas (from 0 to 1). 

Finally, ionised gas that is not subject to ionising radiation will recombine and eventually cool.  We take the characteristic timescale for the recombination to be $(\alpha_{\rm B} n)^{-1}$.

\subsubsection{Implementation}

The main difference between our method and that of \citet{Dale2007b} is the way in which the integrals along the line of sight between the ionising source and a particular gas particle are calculated.   \citet{Dale2007b} used a method similar to that developed by \citet{Kessel-Deynet2000} which approximates the integrals using the density and ionisation state of the closest SPH particles to the line of sight.  Instead, we use SPH interpolation to determine the line of sight integrals using all SPH particles whose smoothing lengths overlap with the line of sight.  This is done by walking the tree structure that is used to determine SPH neighbours and self-gravity.  For example, the second integral in equation \ref{eq:timedependent} becomes
\begin{equation}
\begin{aligned}
\int^{r}_{0} r^{\prime 2} n(r^\prime) [1 - &  H_{\rm II}(r^\prime)] {\rm d}r^\prime =  \\
& \sum_j r_j^2 \chi(x_j,h_j) \frac{2 m_j}{\mu m_{\rm H} h_j^2} (1-H_{{\rm II},j}),
\end{aligned}
\end{equation}
where $\mu$ is the mean molecular weight of the unionised molecular gas, and $m_{\rm H}$ is the mass of a hydrogen atom. The sum is done over all particles that overlap with the line of sight from the ionising source to the particle in question, $r_j$ is the distance from the ionising source to particle $j$, and $m_j$, $h_j$, and $H_{{\rm II},j}$ are the mass, smoothing length, and ionisation fraction of particle $j$, respectively.  The function $\chi(x_j,h_j)$ is the line integral through the SPH kernel function $W(r_j,h_j)$ at an impact parameter of $x_j$ divided by the radius of the smoothing kernel (e.g. $2h_j$).

For each gas particle, we then evolve the ionisation fraction as
\begin{equation}
\begin{aligned}
\frac{{\rm d}H_{\rm II}}{{\rm d}t} = \frac{h^2}{r^2} \left( \frac{Q_{\rm H}}{4 \pi} \right. &  -  \int^{r}_{0} r^{\prime 2} n(r^\prime)^2 \alpha_{\rm B} {\rm d}r^\prime  \\ 
& -  \left. \frac{1}{\delta t} \int^{r}_{0} r^{\prime 2} n(r^\prime) [1-H_{\rm II}(r^\prime)] {\rm d}r^\prime  \right),
\label{eq:evol_ionisation}
\end{aligned}
\end{equation}
as long as this quantity is positive.  The factor $h^2/r^2$ accounts for the fact that the ionising flux is distributed over $4\pi$ steradians and only a fraction of this is intercepted by a particular SPH particle.  Although an SPH particle has a total extent given by the SPH kernel (which, for the standard cubic spline has a radius of $2h$), these regions overlap so, although the effective solid angle subtended by the particle should scale as $h^2$, the coefficient is unclear.  Empirically we find that $h^2$ gives good results (see Appendix \ref{app:mass_resolution}).

We treat photoionisation from multiple sources by summing the positive contributions to fractional change in $H_{\rm II}$ from all ionising sink particles. For gas particles for which this sum is zero, we set
\begin{equation}
\frac{{\rm d}H_{\rm II}}{{\rm d}t} = -  n\alpha_{\rm B} H_{\rm II}^2.
\label{eq:evol_recom}
\end{equation}
This allows ionised gas that does not receive ionising flux to recombine.

The columns for every particle-to-sink line of sight below a distance threshold are calculated, and $H_{\rm II}$ values are evolved, on the same individual timesteps as all other derivatives. We set the distance threshold to 100 pc which reduces the computational expense by a factor of a few to ten, depending on the size of a simulation and the number of ionising sinks. We run one simulation, SR\_los, without a limit to determine the validity of this 100pc limit. In the IA\_ion run we use a limit of 500 pc to reflect the much larger distances between star formation locations in an inter-arm region.

\begin{figure*}
  \includegraphics[width=\textwidth]{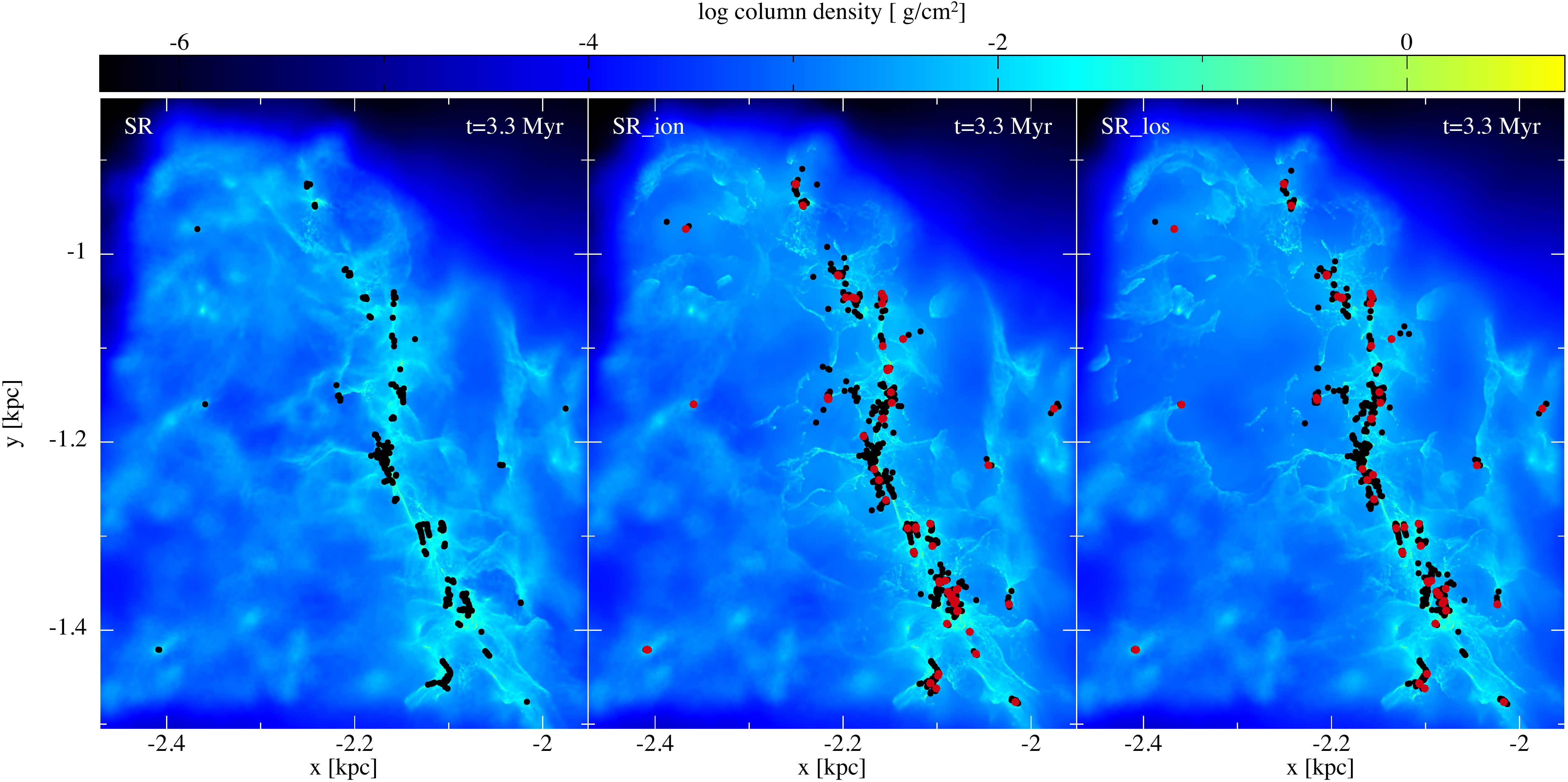}
  \par\bigskip
  \includegraphics[width=\textwidth]{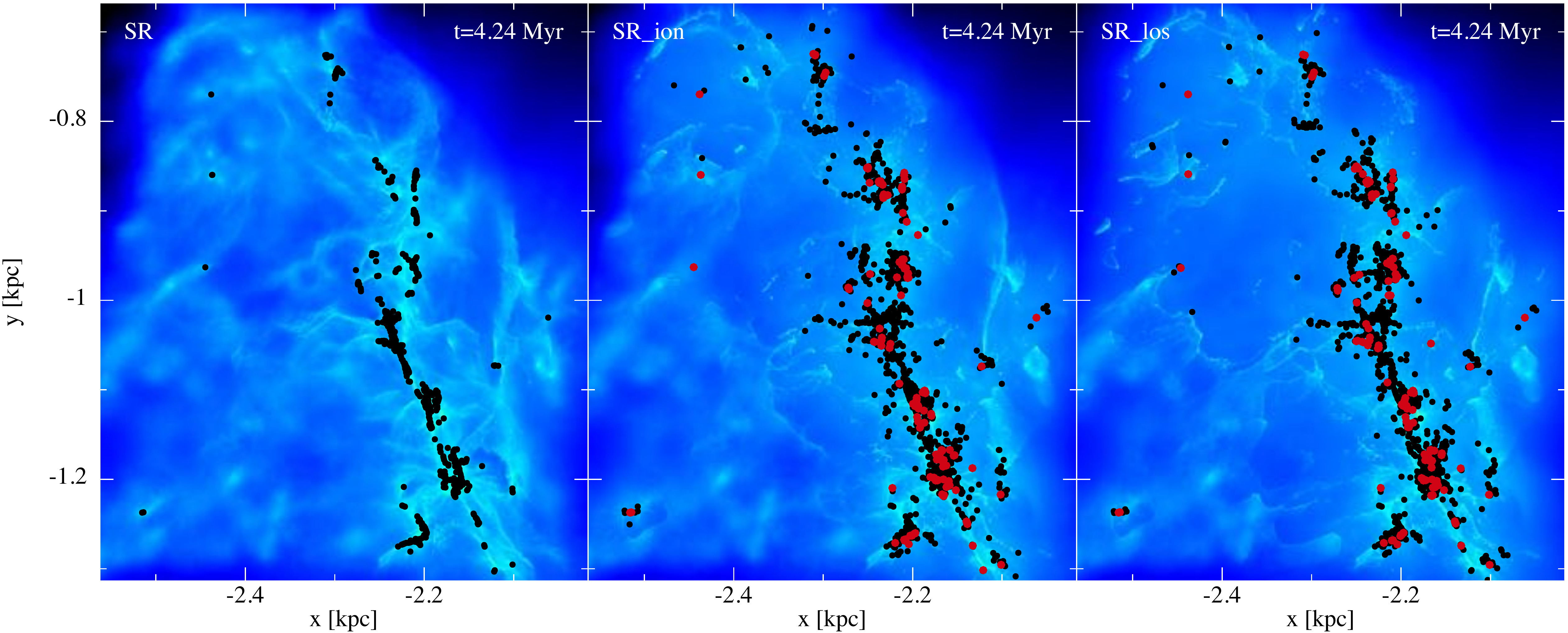}
  \caption{The gas surface density is shown for the standard resolution runs at 3.3 Myr (top row) and 4.2 Myr (bottom row), with no feedback (left), ionising feedback limited to a range of 100 pc (middle), and with no distance limit (right). Ionising sinks are distinguished by the larger, red dots. The runs with ionisation show sharper features in the gas, and more sink particles, including sink particles at locations which otherwise do not show star formation without ionisation at the same timeframe. Long range features are more visible in the right hand panel compared to the middle.} 
  \label{fig:wholesim}
\end{figure*}

\subsection{Choice of SPH mass resolution}
\label{sec:resolution}

We perform multiple single source tests of the photoionisation algorithm at a variety of resolutions in order to determine a lower limit to mass resolution for these SPH simulations. Three factors determine whether an ionised source produces an adequately resolved HII region: the SPH mass resolution, the ambient density around the sink particle and the ionising flux of the source(s). If the resolution is too low HII region sizes are underestimated and the shock wave is broader and less well defined. We present detailed analyses of our tests in Appendix \ref{app:mass_resolution}. In summary, in our simulations at 1 M$_\odot$ per SPH particle, 94\% of all ionising photons emitted contribute to HII regions that will reach at least 98\% of the size they would reach if resolved to convergent precision.

\subsection{Simulations}

Table \ref{tab:simulations} summarises all of the simulations presented in this paper. The SR and SR\_ion runs are our fiducial simulations, they are centred on a section of spiral arm of size 0.5 kpc $\times$ 0.5 kpc at a galactic radius of 2 kpc. We compare these to a variety of other simulations each with one key modification to simulation parameters or initial conditions.

We perform three main comparison runs to the fiducial runs using identical initial conditions. We typically assume that all mass that ends up in sink particles forms stars, however, in the SR\_50\% run, we only allow half of this gas to form stars. This should lead to half as much stellar mass and, therefore, half as much ionising radiation, although this is dependent on the sample distribution. We use this run to identify the relative effects of varying levels of photoionisation. As discussed in Section \ref{sec:ionmethod} we cap all lines of sight at 100 pc, therefore ignoring long range ionisation, but in the SR\_los run we relax this condition. \citet{Emerick2018} suggest that such long range ionisation can be vital to drive outflows in dwarf galaxies. Run SR\_los is used to assess the importance of long range ionisation in this work. We also use an alternative sample of stars in SR\_alt, this just has a different order of massive star creation.

The HR and HR\_ion runs use the fiducial initial conditions but with three times the mass resolution. These runs are a means to check the work on resolution choice discussed in Section \ref{sec:resolution} and Appendix \ref{app:mass_resolution} in anisotropic initial conditions with multiple ionising sources. The IA and IA\_ion runs use initial conditions from a separate inter-arm region. Due to lower densities and fewer sink particles forming we are able simulate a larger region of 1.0 kpc$\times$ 1.0 kpc (see Figure \ref{fig:ICs}). These models investigate how the impact of photoionisation varies inside and outside spiral arms.

We also vary the accretion radius of our sink particles. Different radii vary the scale over which sinks form. Smaller radii will lead to a larger number of smaller sinks. There is a computational benefit to fewer sink particles but it also decreases the resolution on which cluster behaviour can be analysed. Modifying sink parameters can also change the total sink mass in simulations, we aim to get an idea of the scale of these differences. Our standard sink accretion radius is 0.78 pc whilst the SR\_2 simulations use 0.45 pc.

\section{Results}
\label{sec:results}

\begin{figure*}
  \includegraphics[width=0.95\textwidth]{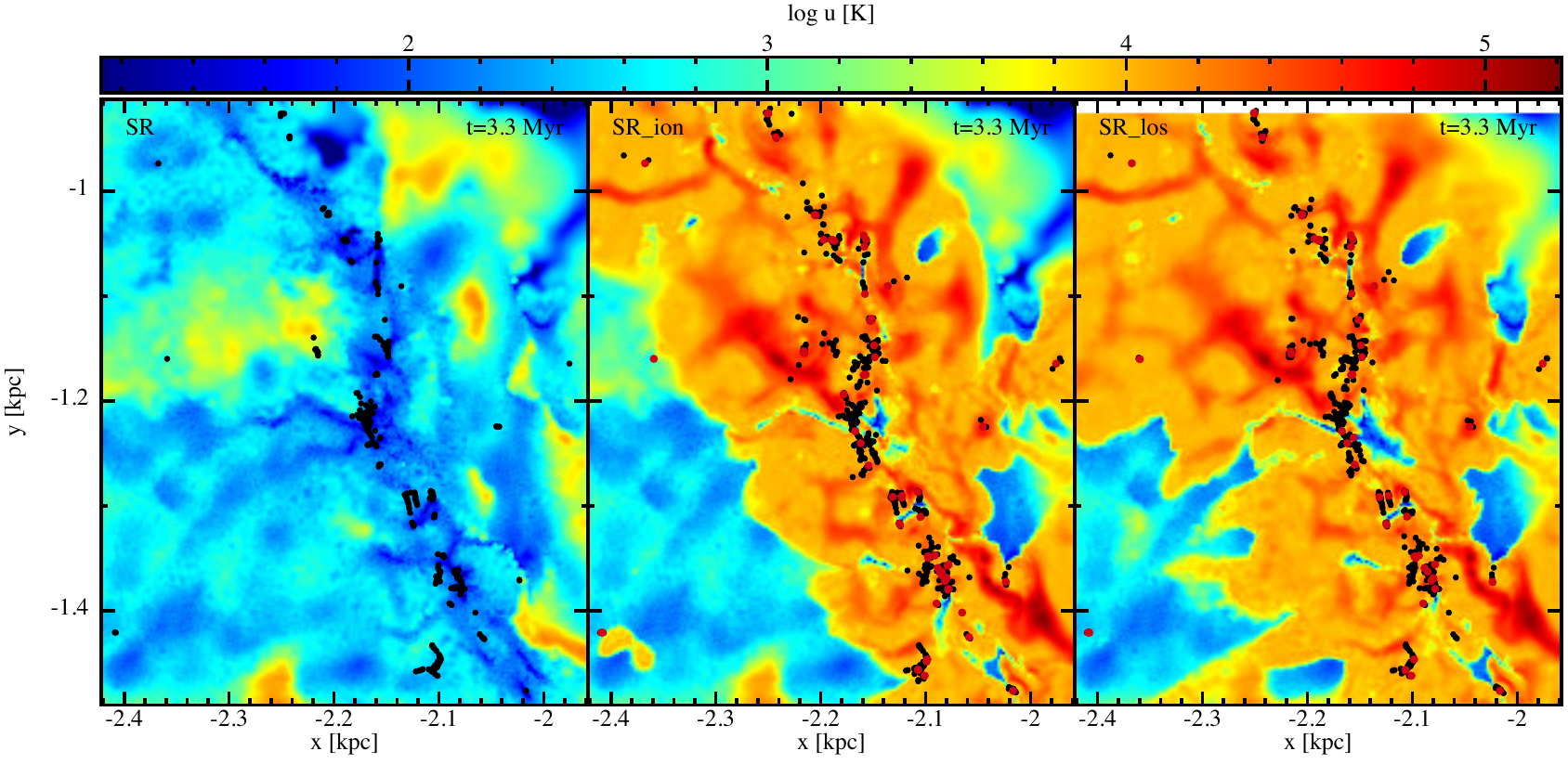}
  \caption{We show a cross-section slice at z=0 of temperatures after 3.3 Myr for the same three simulations as Figure \ref{fig:wholesim}.}
  \label{fig:temperatures}
\end{figure*}

\begin{figure*}
  \includegraphics[width=0.95\textwidth]{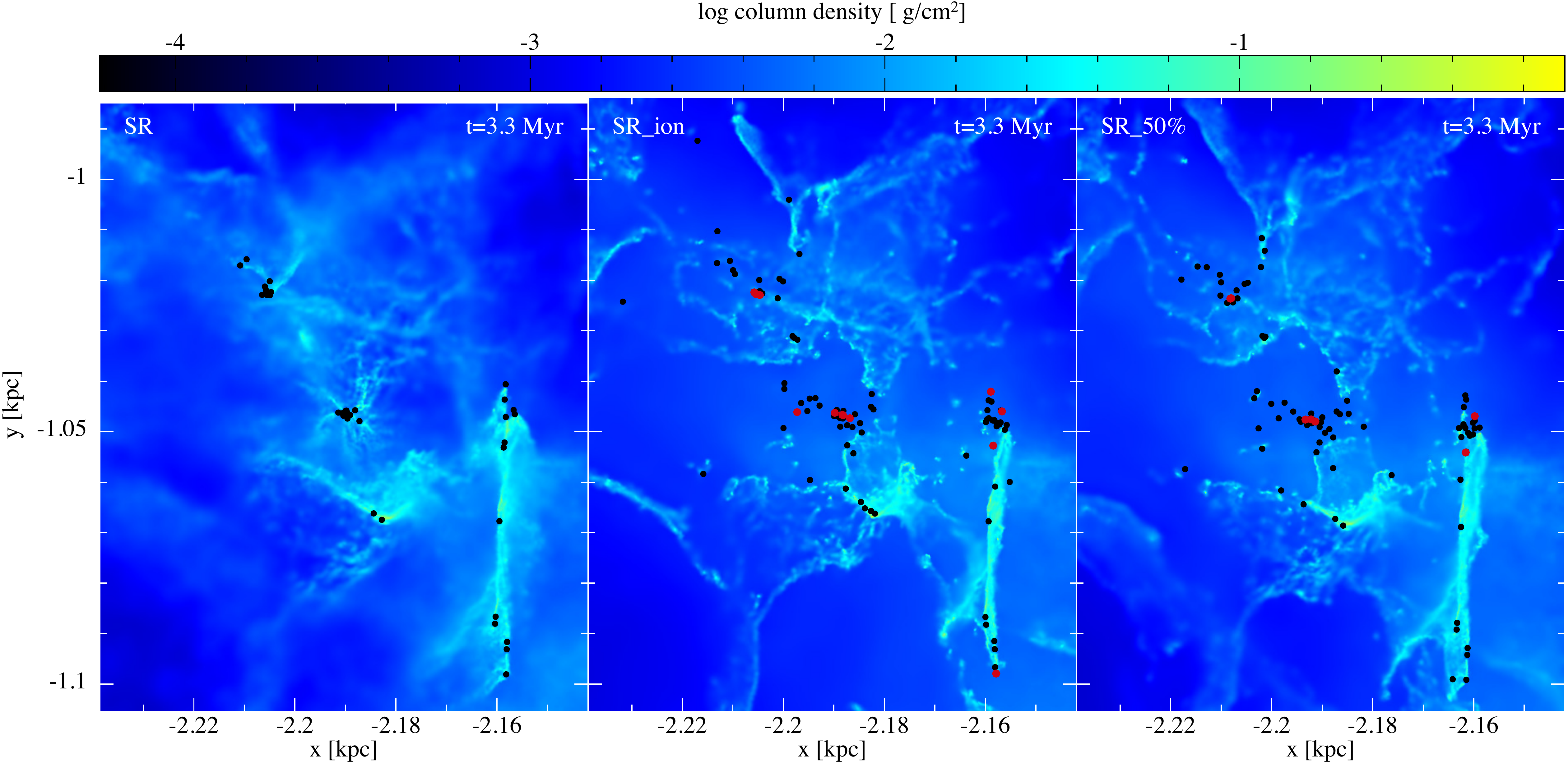}
  \caption{A sub-region ($\approx$ 100 pc\textsuperscript{2}) is displayed where the cluster formation is heavily affected by ionisation. The left and middle panel show the same runs as Figure \ref{fig:wholesim}, but the right panel now shows the SR\_50\% run.}
  \label{fig:ionregion}
\end{figure*}

In this section we include a study of the star formation rates in different models (Section \ref{sec:starformation}), the cloud (or clump) distribution (Section \ref{sec:clouds}), the properties of stellar clusters (Section \ref{sec:clusters}) and a comparison with the inter-arm region (Section \ref{sec:inter-arm}), but first we give a comparison of the overall morphology of the different simulations. We mostly concentrate on comparisons between the SR run with and without ionisation. Figure \ref{fig:ICs} shows the initial conditions for the SR run. For this section of spiral arm, the lower region of the arm is fairly continuous, whereas the upper part has some emptier regions and shells (caused by supernova feedback in the progenitor simulation). In Figure \ref{fig:wholesim} we show simulations of this region at times of 3.3 Myr and 4.24 Myr. The figure shows models without photoionisation in the left panel, with photoionisation in the middle and without an ionisation line of sight limit on the right. The larger red dots represent the ionising sink particles and the smaller black dots represent non-ionising sinks. The gas in the upper half of the spiral arm is visibly dispersed while the lower half appears to be relatively untouched.

In the SR\_los run very large scale ionised cavities are present, especially in the upper half of the region. The largest galactic HII regions which have both measured radii and distances in the WISE catalog \citep{Anderson2014} are tens of pc across. Approximately 10\% of these regions are larger than 10 pc in diameter and $\approx$ 1\% are larger than 50 pc. The size of our largest simulated HII regions is consistent with the largest WISE objects.

Filaments near to ionising sources are more compact and clearly defined where they have been compressed by the ionised gas. We also see a much larger number of sink particles in the runs with photoionisation. This is due to both triggered sink formation which leads to star formation over a wider region of space, and the reduction of accretion onto ionising sinks which leads to a larger number of lower mass sink particles.

The smaller accretion radius simulations (SR\_2) contain a higher number of smaller sink particles, meaning better sink particle resolution. This method leads to lower SFRs, which are closer to observed rates, however, it also increases computational expense. The lower SFR also means that the SR\_2\_ion run has a lower ionising flux than the SR\_ion run. The alternative sample of massive stars (SR\_alt run) also leads to a lower total flux as a result of the stochastic sampling. For the statistical analysis in this paper on star formation (rates, efficiencies, and locations) and cloud properties (mass functions, cloud evolution and virial parameters) the differences between simulations are predominantly governed by the total ionising flux emitted. Therefore, for most of our analysis we choose to focus on the SR\_50\% run, rather than the SR\_2\_ion and SR\_alt runs, since the flux difference between this and the fiducial run is consistent over time and easily quantifiable.

We see the way in which photoionisation modifies the internal energy of the gas in the temperature maps in Figure \ref{fig:temperatures}. Inside the spiral arm only very localised regions are able to remain cool. The lack of cold gas in the upper half of the SR\_ion run emphasises the high impact of photoionisation in this region. The wider extent of heating in the right hand panel (SR\_los model) shows the additional impact of long range radiation.

Figure \ref{fig:ionregion} shows a small part of a region that is heavily affected by photoionisation. In the no feedback case (left panel) we see groups of sink particles lying along, or at, the intersection of filaments. With ionising feedback, the filaments surrounding the star particles are significantly disrupted, and the remaining dense gas appears to be compressed into sharper features. The morphology of these regions is not too dissimilar to previous single cloud simulations of low mass ($\sim10^4$ M$_{\odot}$) clouds \citep{Dale2014,Geen2018}. 

\subsection{Star Formation}
\label{sec:starformation}

\begin{figure}
  \includegraphics[width=\columnwidth]{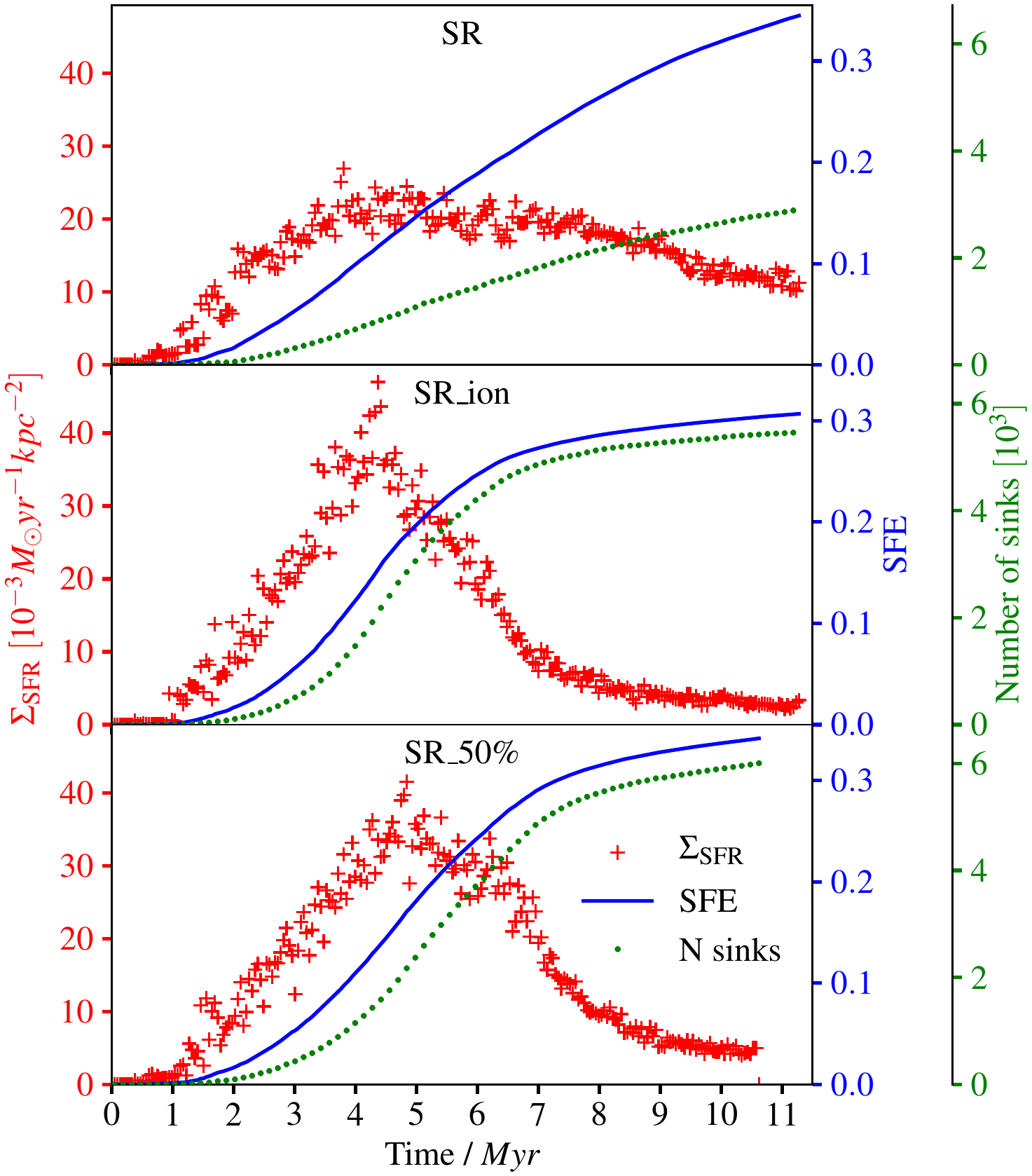}
  \caption{The instantaneous SFR surface density is plotted over time without feedback (top), with photoionisation (middle) and in the SR\_50\% run (bottom) as red crosses. We also plot the absolute star formation efficiency (blue line) and the number of sink particles (green dots).}
  \label{fig:sfr}
\end{figure}

\begin{figure}
  \includegraphics[width=\columnwidth]{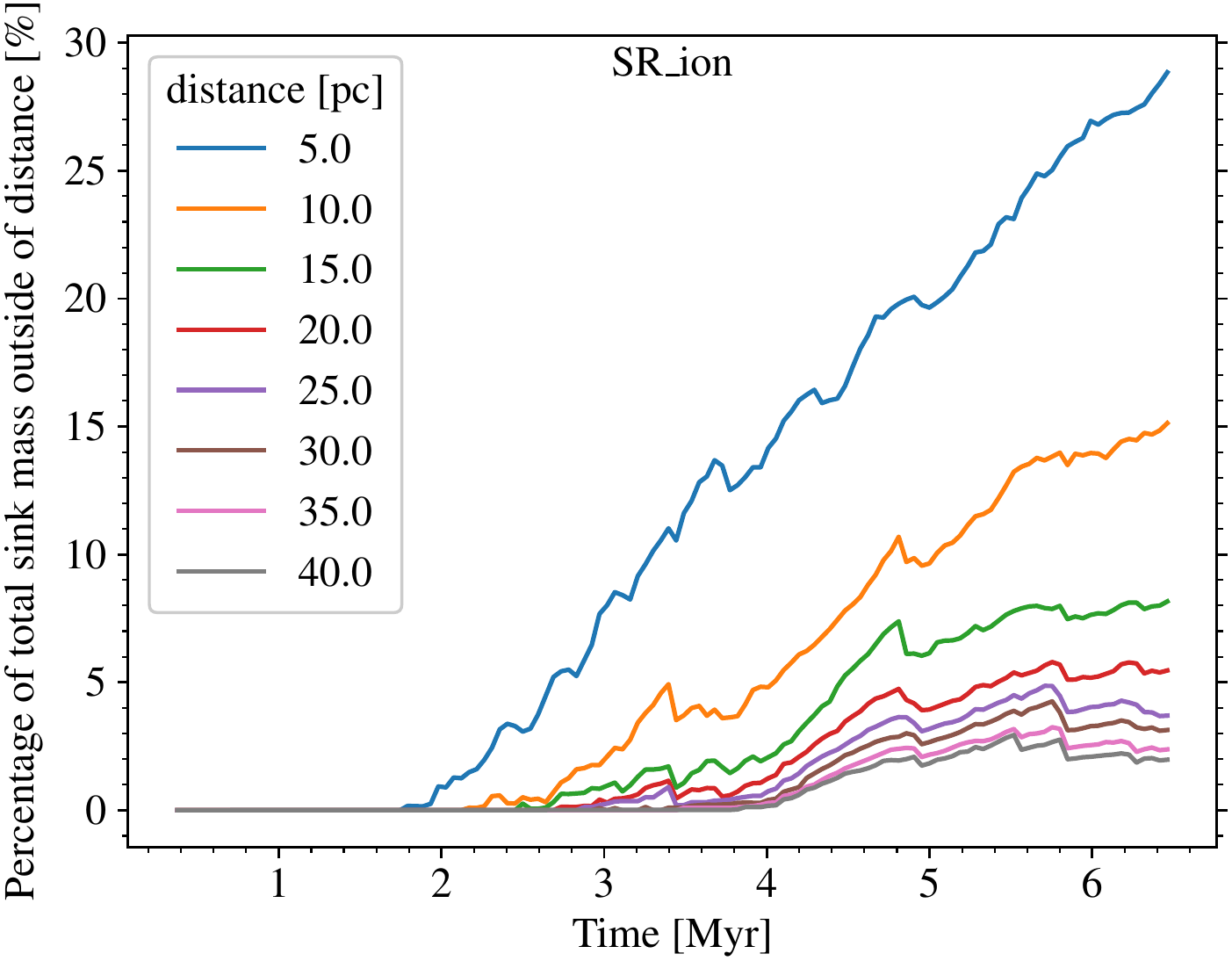}
  \par\bigskip
  \includegraphics[width=\columnwidth]{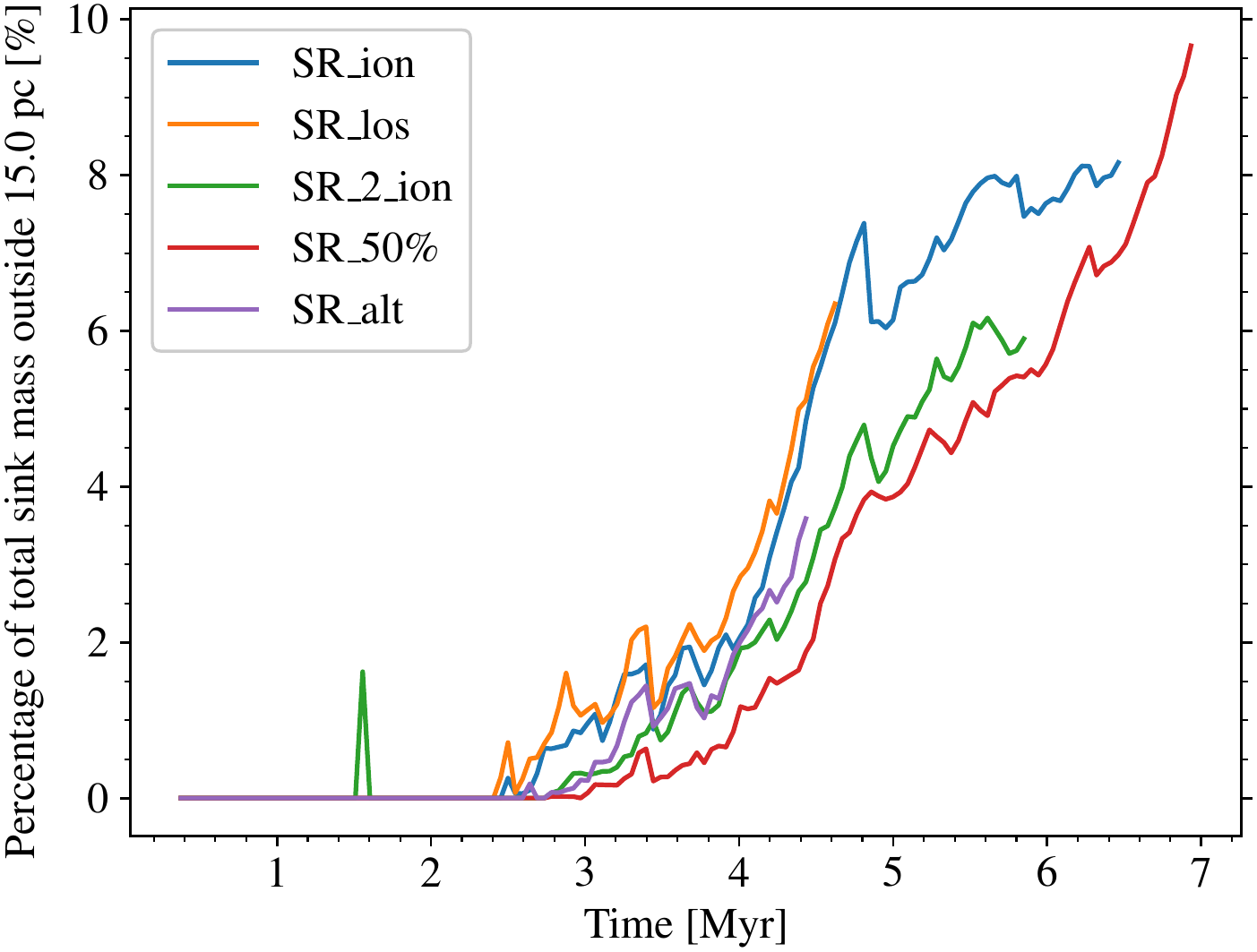}
  \caption{Top panel: the percentage of mass in sinks in the SR\_ion run that is further than a range of distances from the nearest sink particle in the SR run. Bottom panel: the top panel's 15 pc distance plot now compared to several runs with photoionisation.}
  \label{fig:sinkdist}
\end{figure}

\begin{figure}
  \includegraphics[width=0.98\columnwidth]{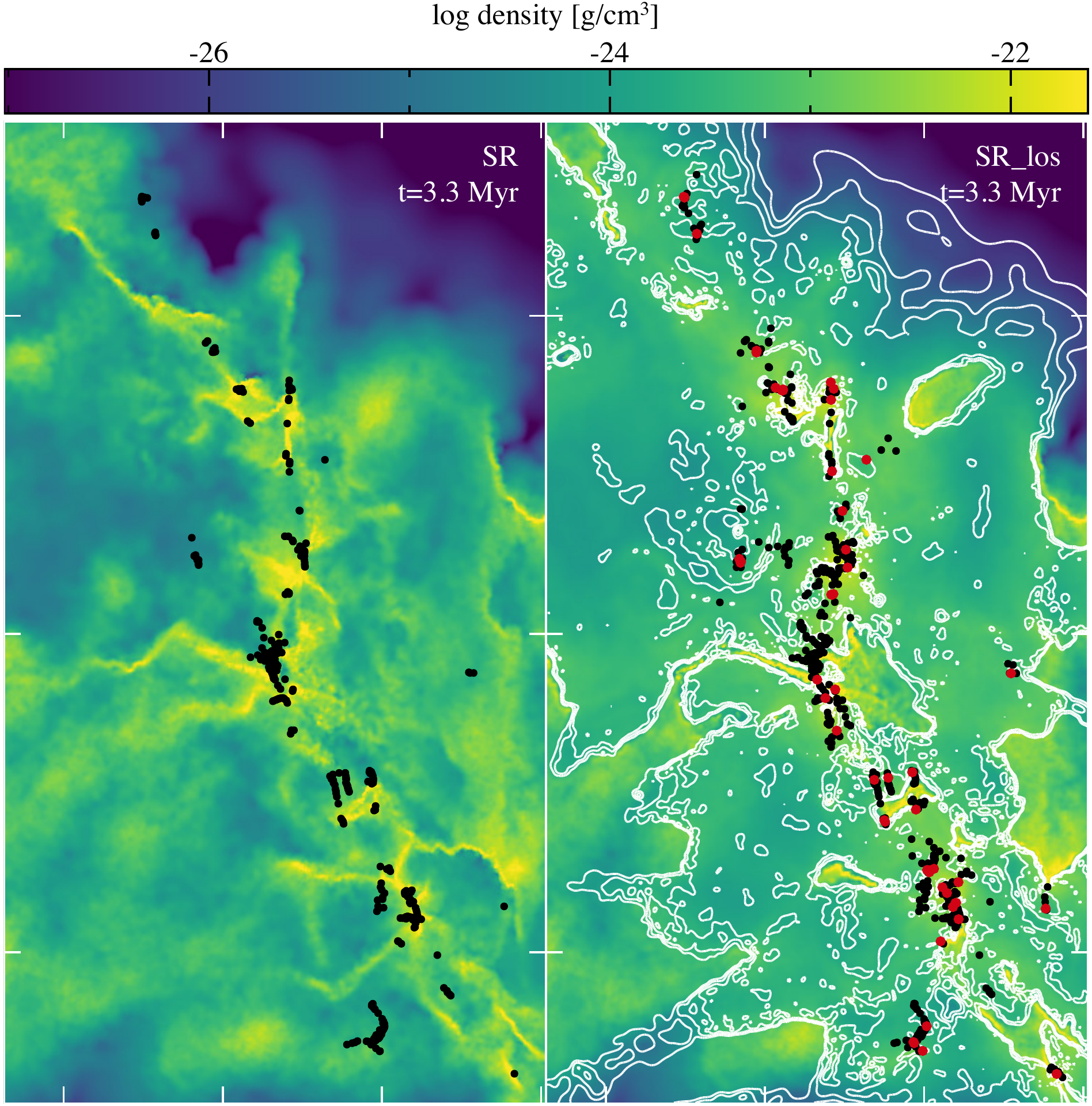}
  \includegraphics[width=0.98\columnwidth]{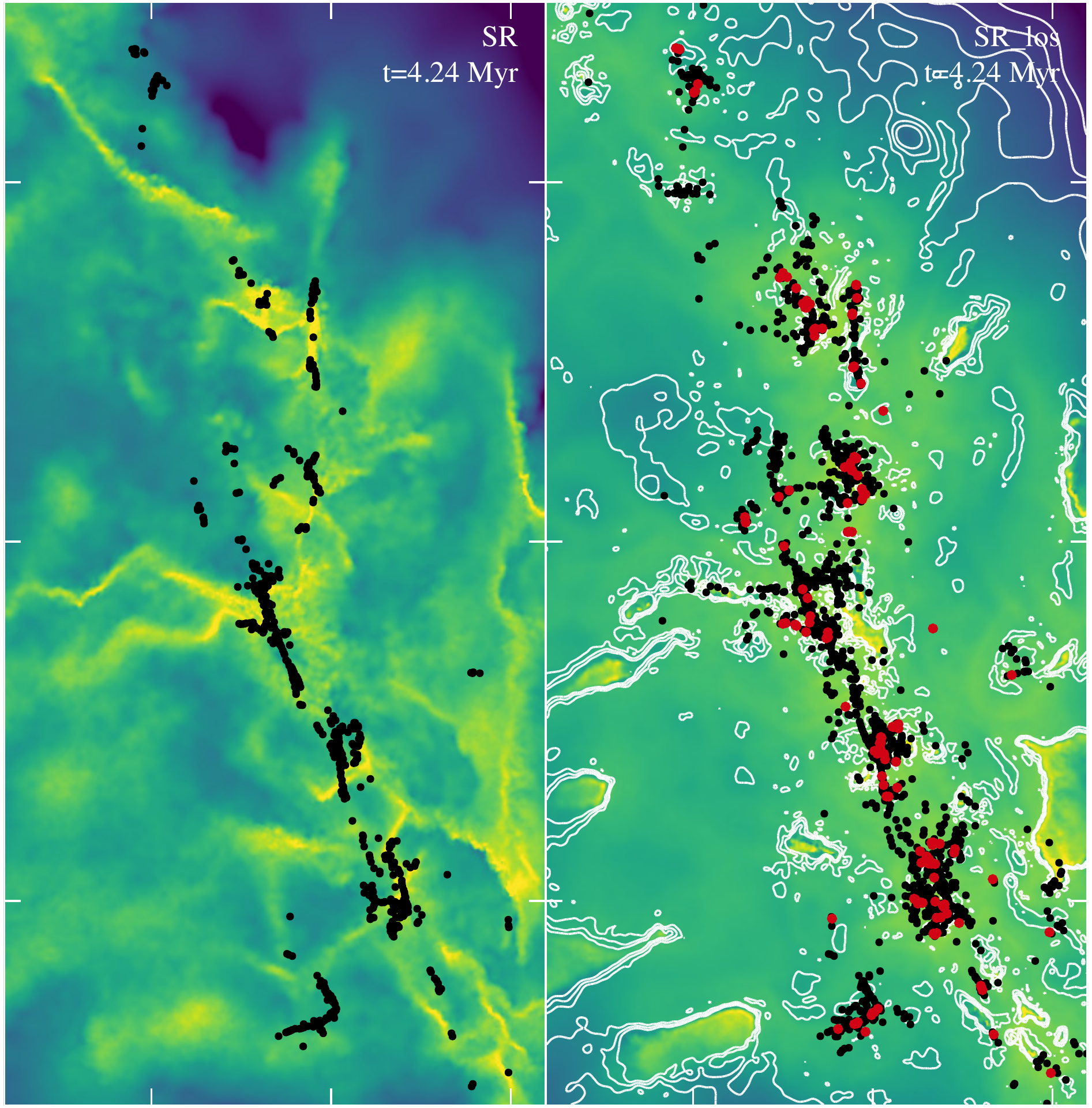}
  \caption{Left panels: run with no photoionisation (SR). Right panels: No range limit photoionisation run (SR\_los) with ionised gas fraction overlaid in contours. The top and bottom panels are at 3.3 Myr and 4.24 Myr respectively. The contours in the top right panel allow for the identifcation of pockets of dense gas that are resisting ionisation. Comparison with the bottom panels shows that when dense pockets are compressed by HII regions from multiple sides triggerred star formation is very likely to occur within a Myr. Gas that only has an HII region on one side may or may not undergo more modest triggered star formation on this timescale. The cross sections are shown about z=0.}
  \label{fig:contour}
\end{figure}

\begin{figure}
  \includegraphics[width=\columnwidth]{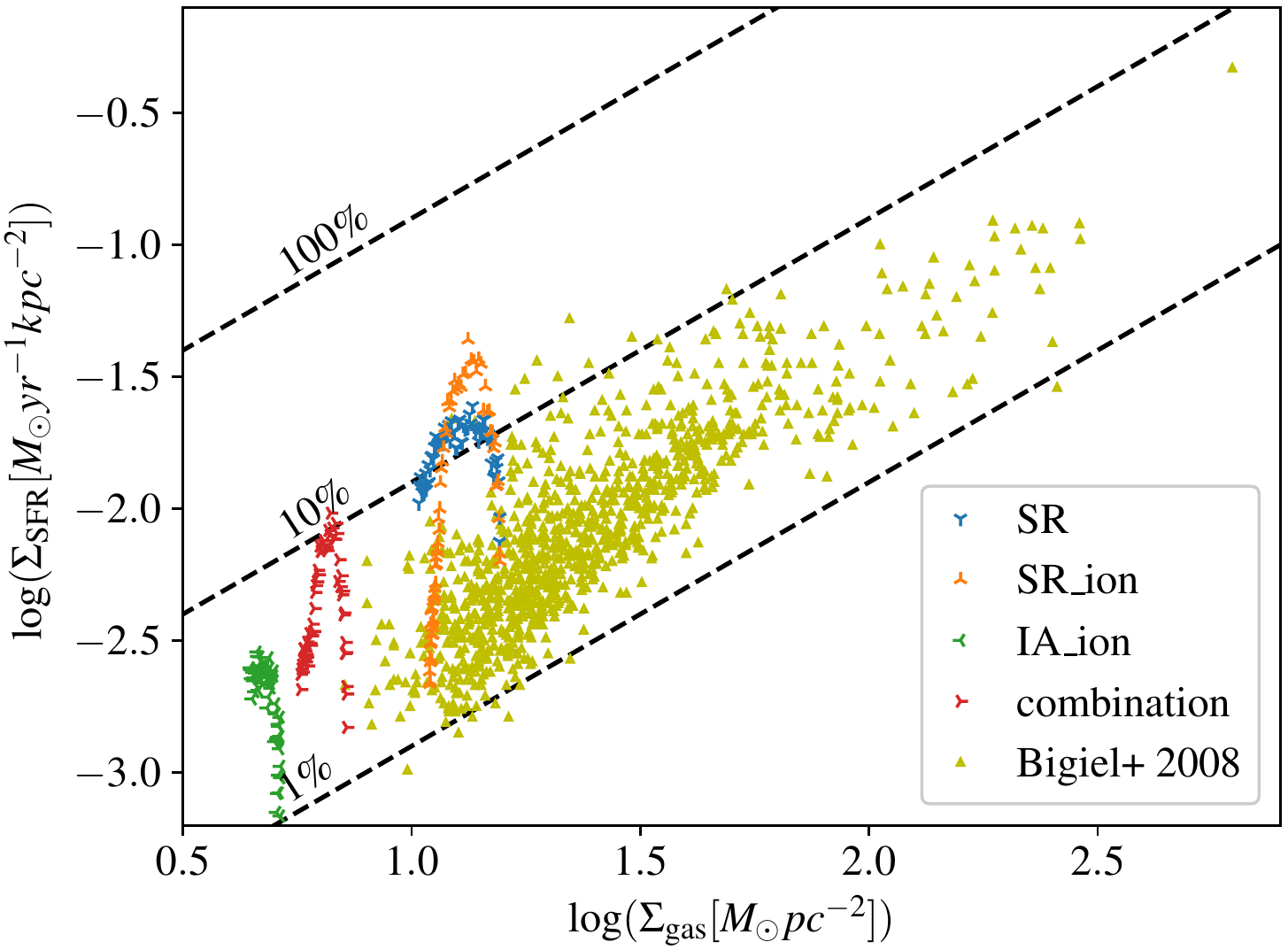}
  \caption{The relation between SFR surface density ($\Sigma_{\mathrm{SFR}}$) and total gas surface density ($\Sigma_{\mathrm{gas}}$) is plotted for our simulations and using observational data from \citet{Bigiel2008}. Multiple points plotted for our simulations represent the time evolution of the region. We plot the spiral arm (with and without feedback), an inter-arm region and a weighted combination of the two such that $\Sigma_{gas}$ is equal to the entire galaxy our initial conditions are extracted from. The diagonal dotted lines, also from \citet{Bigiel2008}, give the absolute star formation efficiency over a $10^8$ year period.}
  \label{fig:kenschmidt}
\end{figure}

In Figure \ref{fig:sfr} we compare the instantaneous star formation rate surface density ($\Sigma_{\mathrm{SFR}}$), absolute star formation efficiency (SFE) and the number of sink particles between the SR, SR\_ion and SR\_50\% runs. We calculate the instantaneous $\Sigma_{\mathrm{SFR}}$ as 
\begin{equation}
\Sigma_{\mathrm{SFR}}(t_n) = \frac{M_*(t_n)-M_*(t_{n-1})}{A\ (t_n-t_{n-1})},
\end{equation}
\noindent
where $t_n$ is the elapsed time at a given timestep, $M_*{(t_n)}$ is the total sink mass in the simulation at a given time and $A$ is the surface area of the initial conditions (which we treat as constant in time) for the simulation in the x-y plane. We calculate the absolute SFE as
\begin{equation}
\mathrm{SFE}(t_n) = \frac{M_*(t_n)}{M_*(t_n)+M_{\rm gas}(t_n)},
\end{equation}
\noindent
where $M_{\rm gas}$ is the total gas mass in the simulation.

As seen in Figure \ref{fig:sfr}, the SFRs in the SR and SR\_ion runs are comparable for the first 2.5 Myr, after which time the star formation in the SR run begins to approach a steady state. However, the SR\_ion run continues to accelerate until 4 Myr, peaking at double the rate, after which the SFR rapidly drops. There is very little difference in the peak $\Sigma_{\mathrm{SFR}}$ between the SR\_ion and SR\_50\% runs, the gradient, however, both before and after the peak, appears to be dependent on the total photoionising flux in the simulation. This trend is also seen in the SR\_2\_ion and the SR\_alt runs. 

We discuss the differences for the inter-arm initial conditions (IA) in Section \ref{sec:inter-arm}. We also see this spike in star formation in the SR\_2\_ion run suggesting it is not a numerical effect resulting from the choice of sink radius.

\begin{figure}
  \includegraphics[width=\columnwidth]{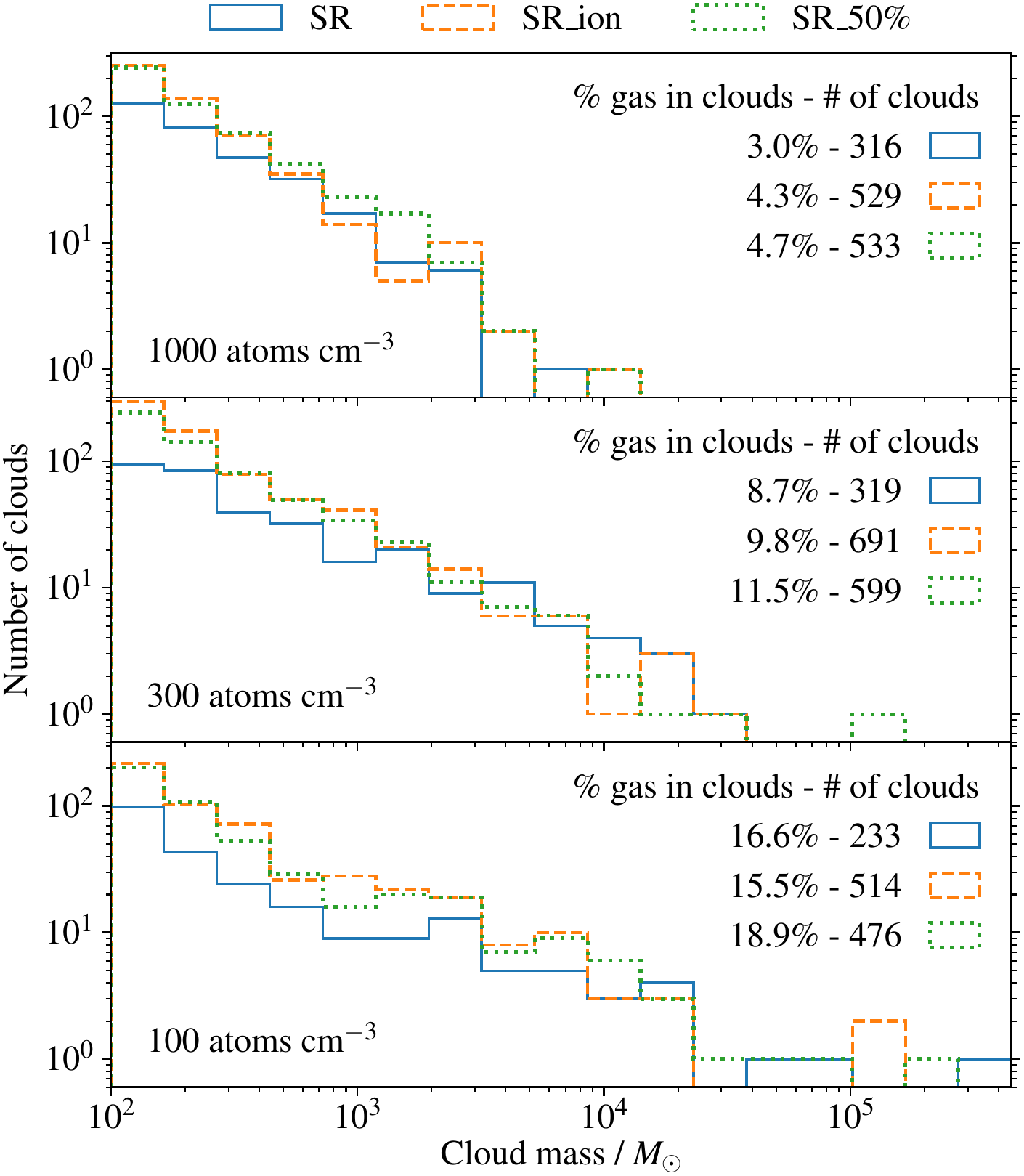}
  \caption{Cloud mass functions are shown for three cloud densities for the SR runs (continuous blue), the SR\_ion run (dashed orange), and the SR\_50\% run (dotted green). The clouds are defined using a friends of friends algorithm with a maximum particle separation criteria of 0.24 pc, 0.37 pc and 0.55 pc (top to bottom), and the median number density of clouds for each of these criteria is $\approx$ 100, 300 and 1000 cm$^{-3}$ respectively.  This is a snapshot at 4.24 Myr chosen, since this roughly corresponds to the peak in the difference in the number of clouds between the SR and SR\_ion runs. We also quote the fraction of all gas in clouds and the number of clouds for each case.}
  \label{fig:massfn}
\end{figure}

\begin{figure}
  \includegraphics[width=\columnwidth]{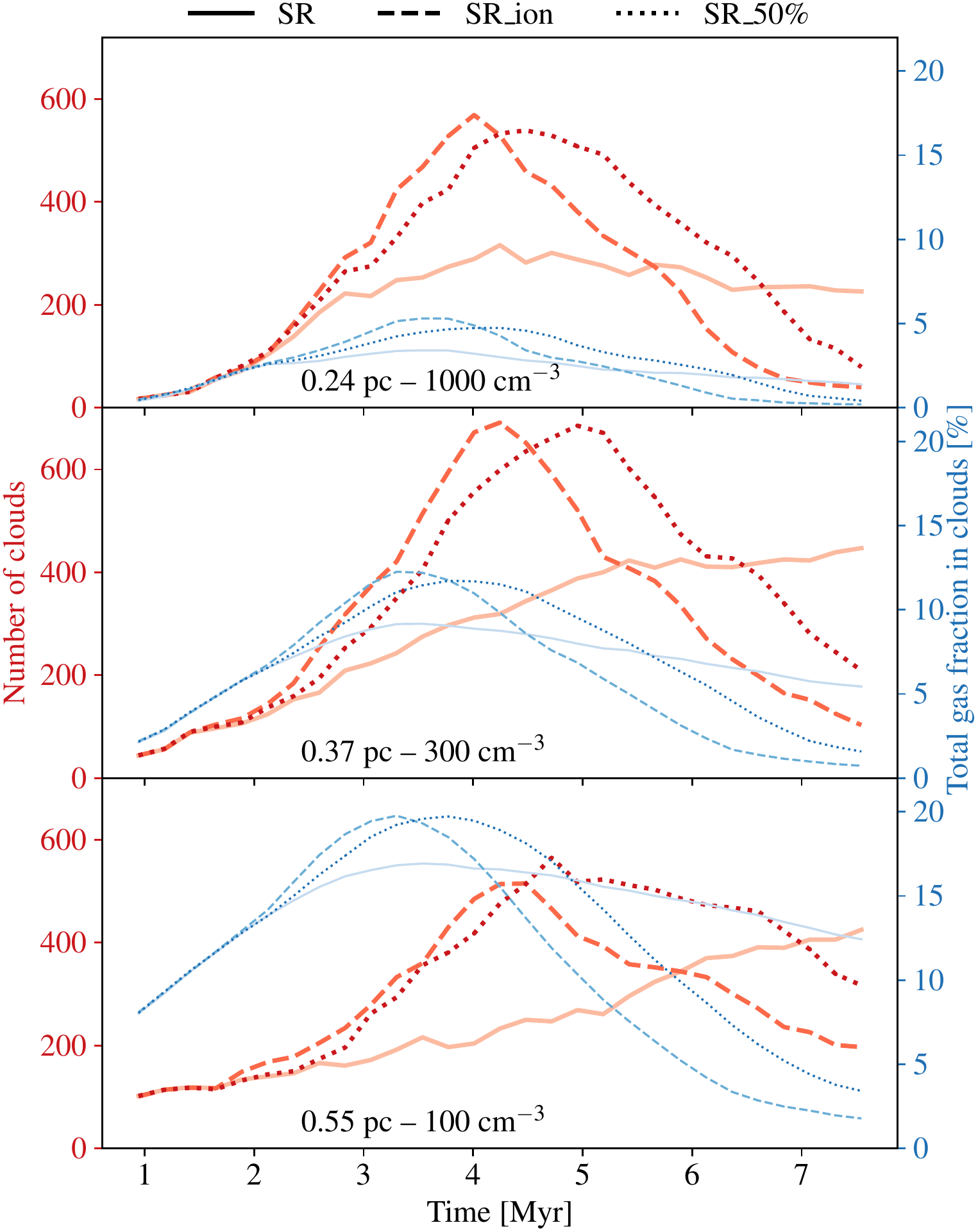}
  \caption{The evolution of the number of clouds (thicker red) and the gas fraction contained in them (thinner blue) is shown with time. We present the same runs and use identical criteria to define the clouds as in Figure \ref{fig:massfn}.}.
  \label{fig:cloud_evol}
\end{figure}

\begin{figure*}
  \centering
  \includegraphics[width=\textwidth]{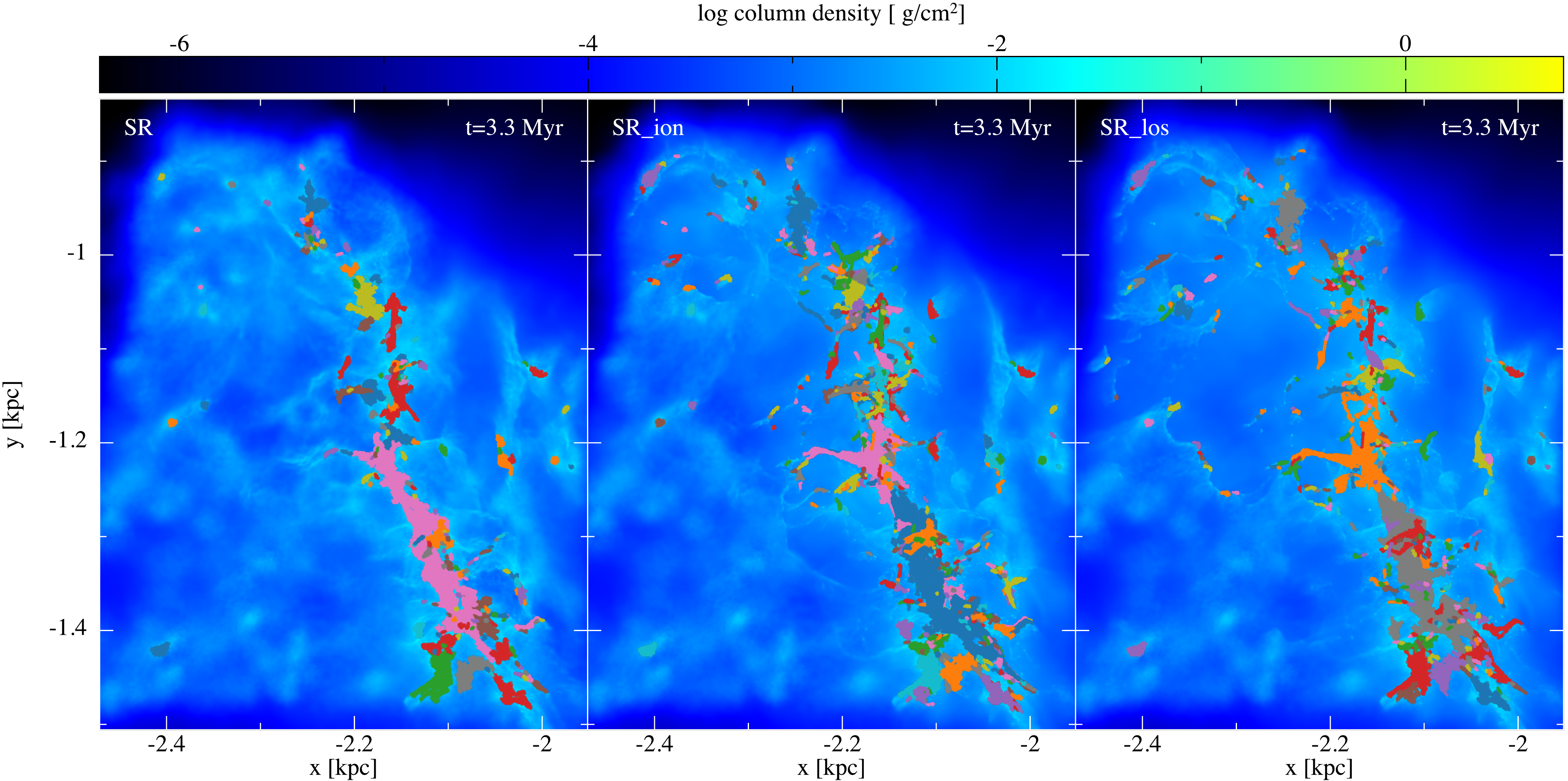}
  \par\bigskip
  \includegraphics[width=\textwidth]{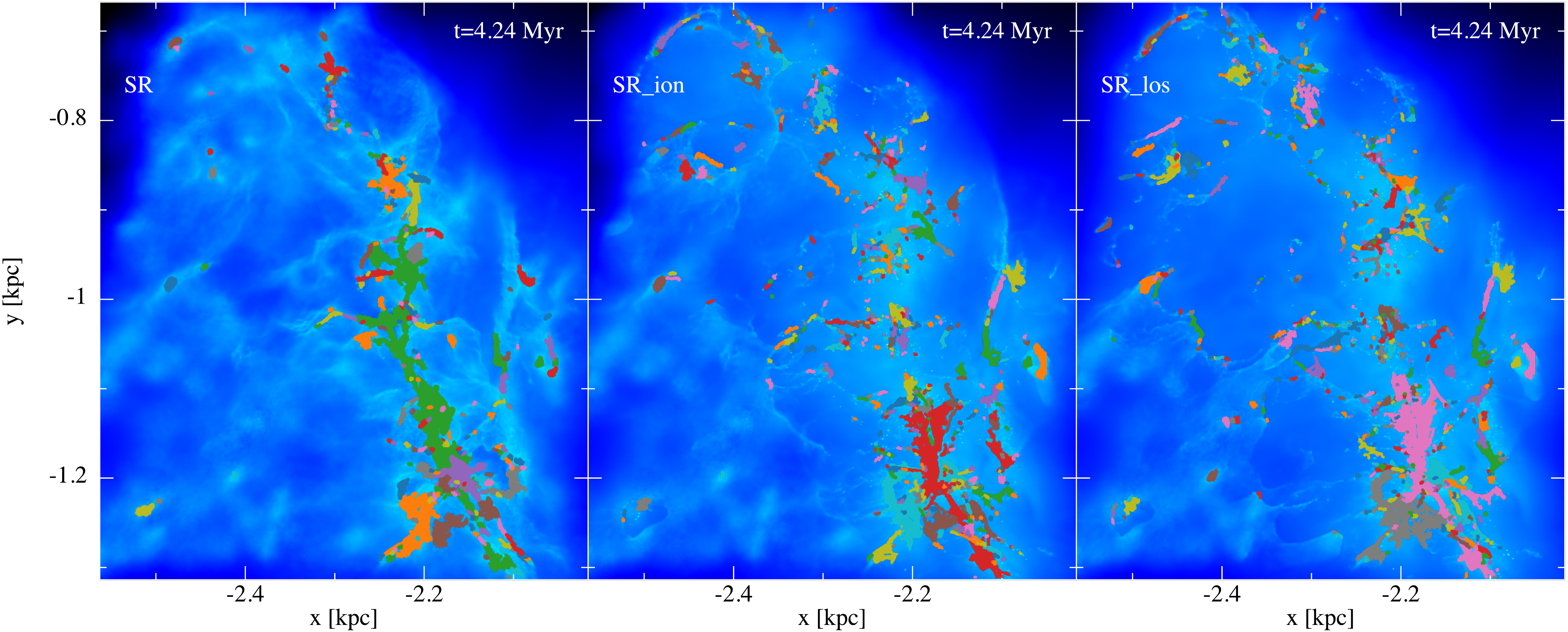}
  \caption{As in Figure \ref{fig:wholesim}, we show the simulations with no feedback (left), SR\_ion (centre), and SR\_los (right), but with clouds of median number density $\approx$ 100 cm$^{-3}$ over-plotted. These clouds were identified using a friends of friends algorithm with a nearest neighbour distance of 0.55 pc.}
  \label{fig:clump_map}
\end{figure*}

\begin{figure}
  \includegraphics[width=\columnwidth]{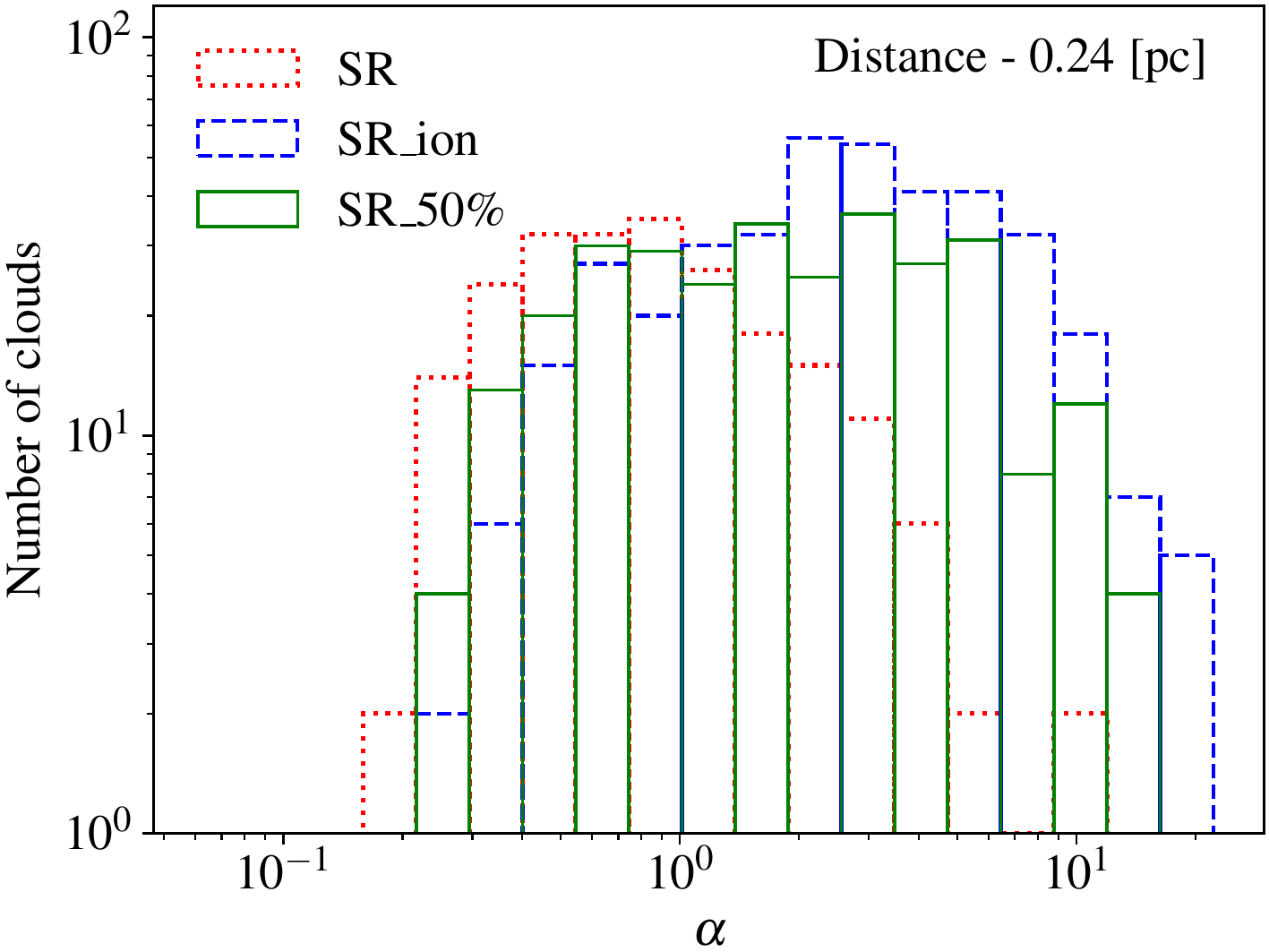}
  \par\bigskip
  \includegraphics[width=\columnwidth]{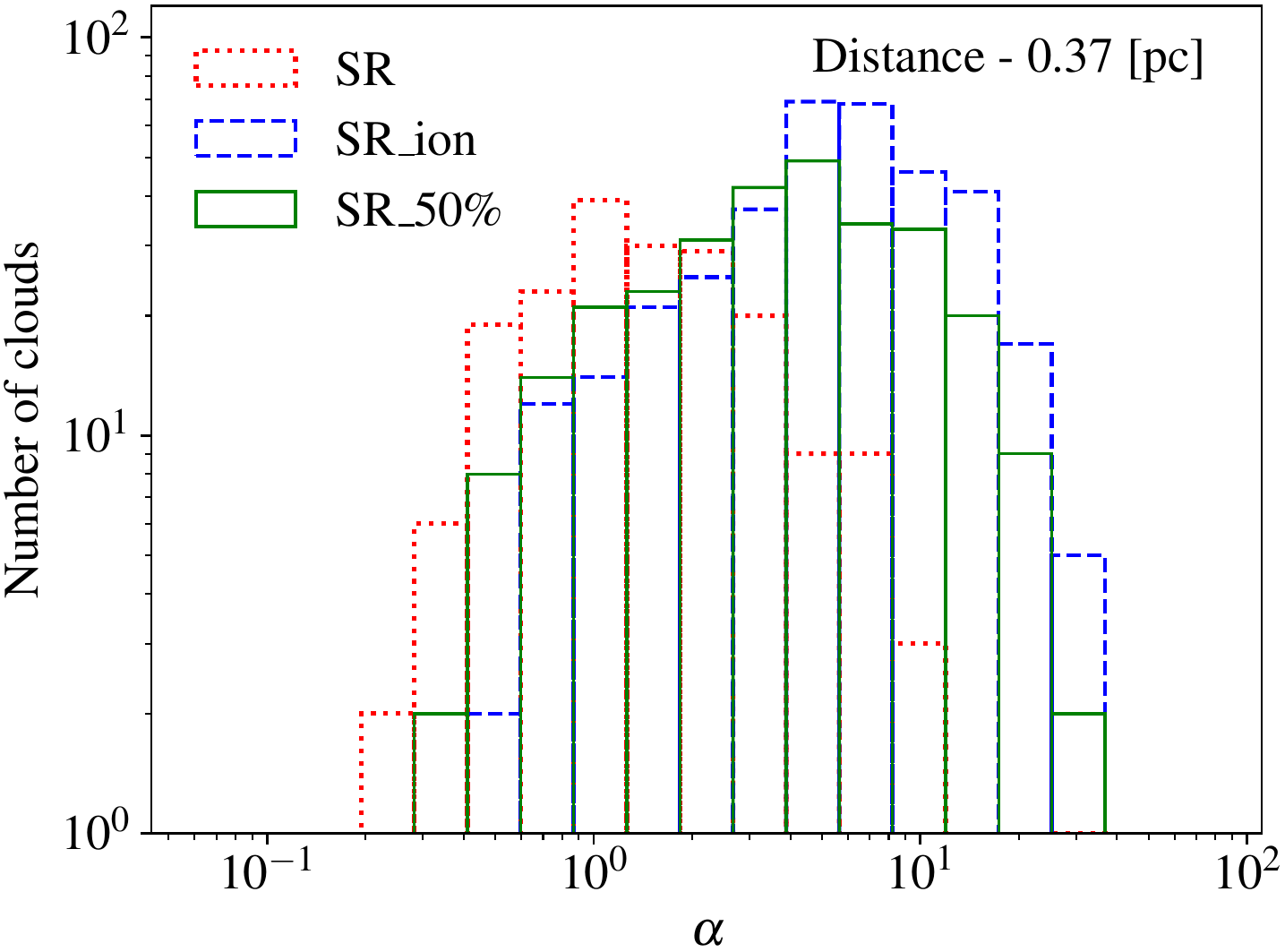}
  \par\bigskip
  \includegraphics[width=\columnwidth]{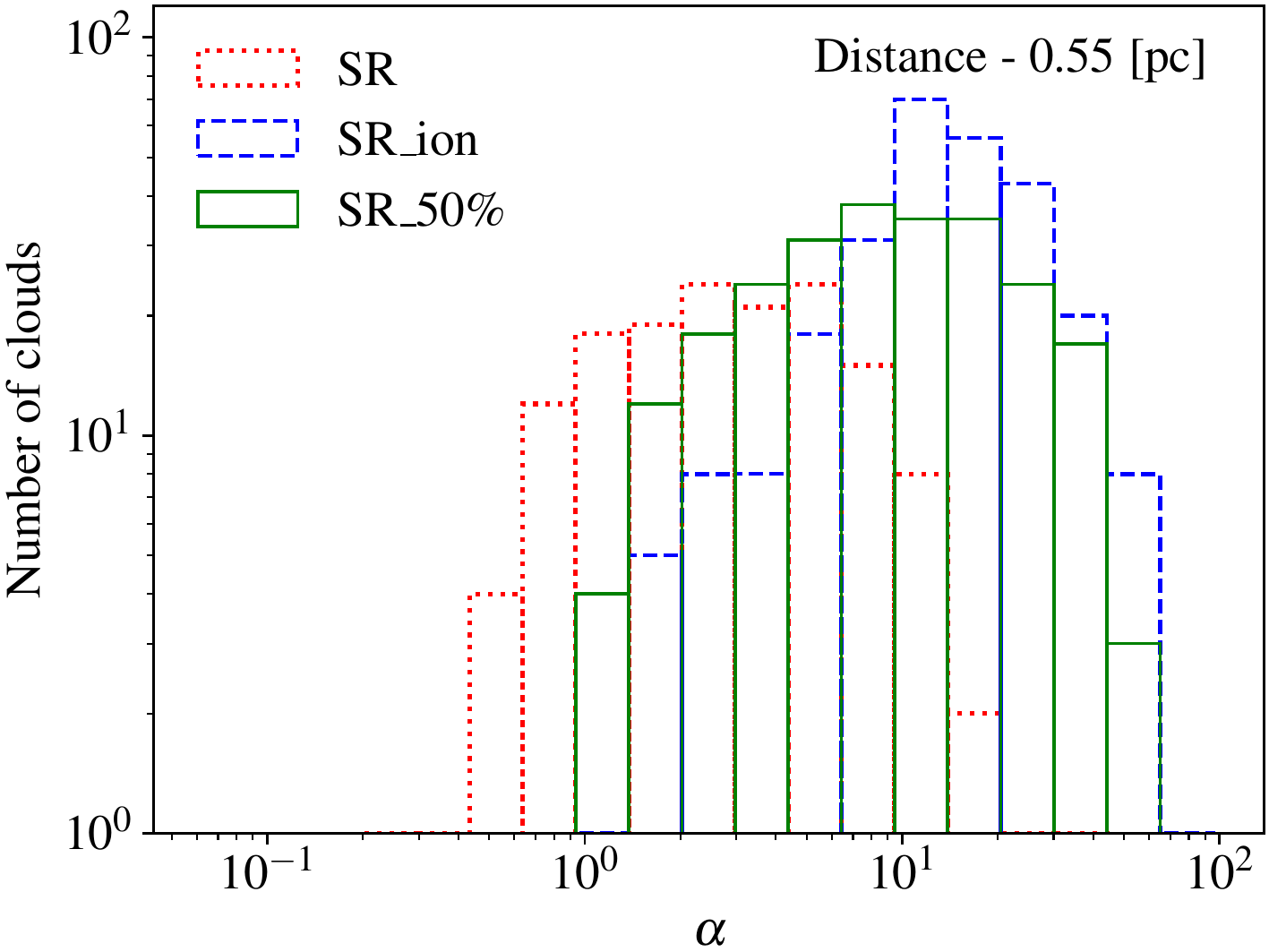}
  \caption{The distribution of the kinetic to gravitational potential energy ratio is shown for all clouds of more than 100 particles (100 M$_\odot$ in these cases). To be considered part of a cloud, a particle must have a number density of at least 50 cm$^{-3}$ and be within the minimum distance of at least one other particle within the cloud. Top, middle and bottom panels correspond to the panels in Figures \ref{fig:massfn} and \ref{fig:cloud_evol}. The fiducial runs with and without photoionisiation are plotted along with the 50\% SFE run.} 
  \label{fig:virial}
\end{figure}

\begin{figure}
  \includegraphics[width=\columnwidth]{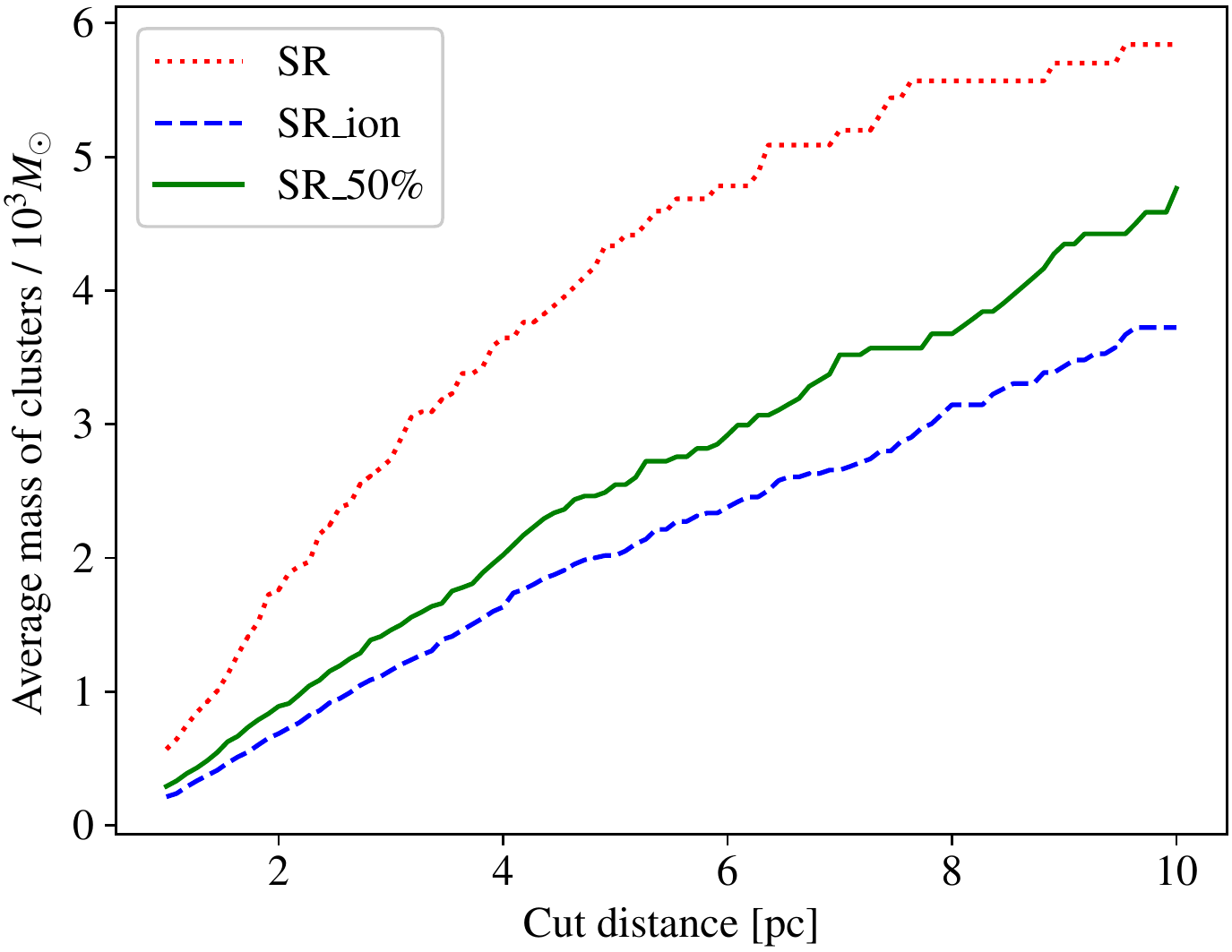}
  \par\bigskip
  \includegraphics[width=\columnwidth]{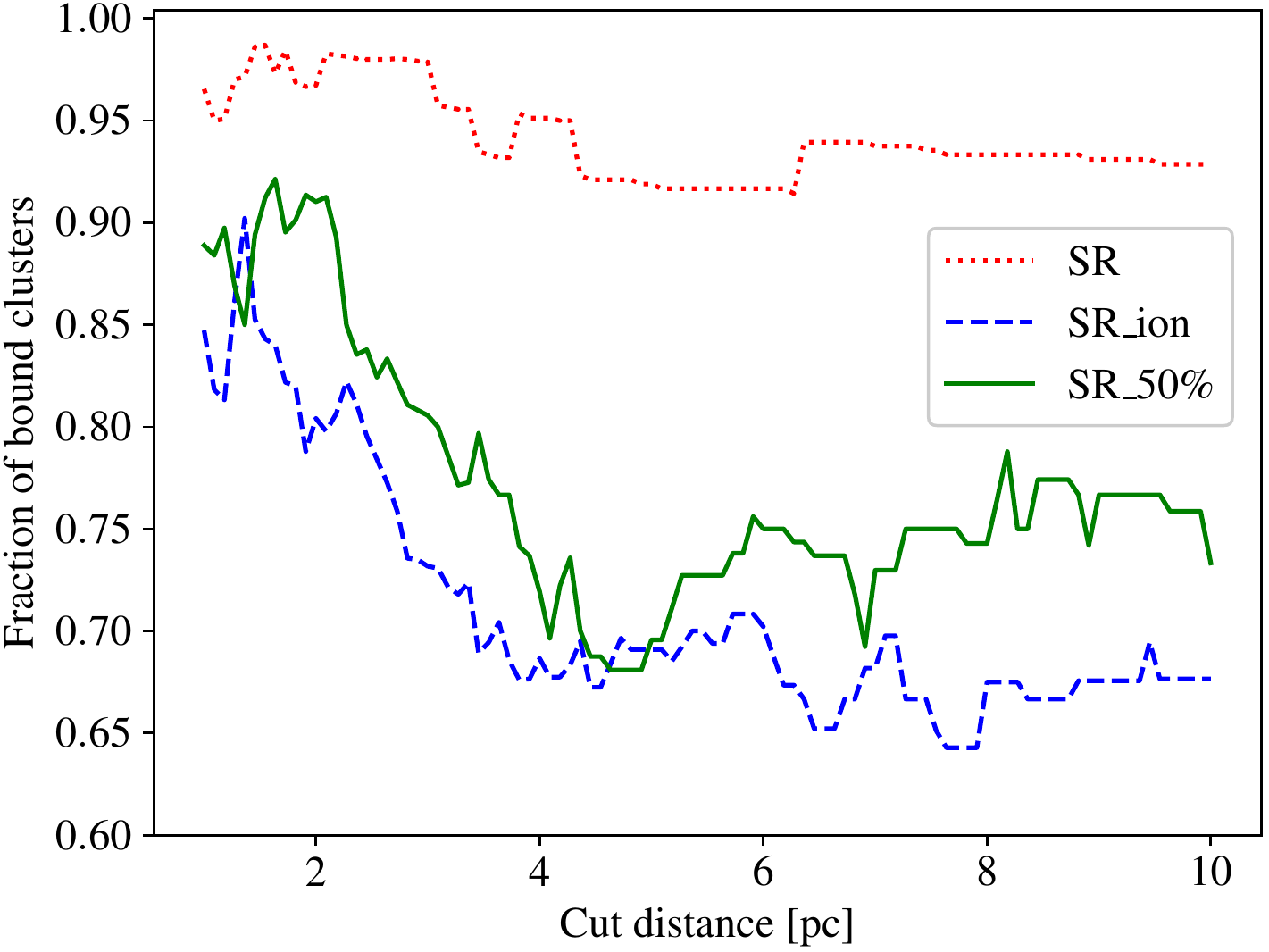}
  \caption{The properties of clusters are shown versus the cut-off distance we take for the minimum spanning tree (which sets the extent of the cluster) at a time of 3.3 Myr. We show the average cluster mass in the upper panel and the fraction of clusters that are bound in the lower panel.}
  \label{fig:mass_bound_clusters}
\end{figure}

\begin{figure*}
  \includegraphics[width=0.95\textwidth]{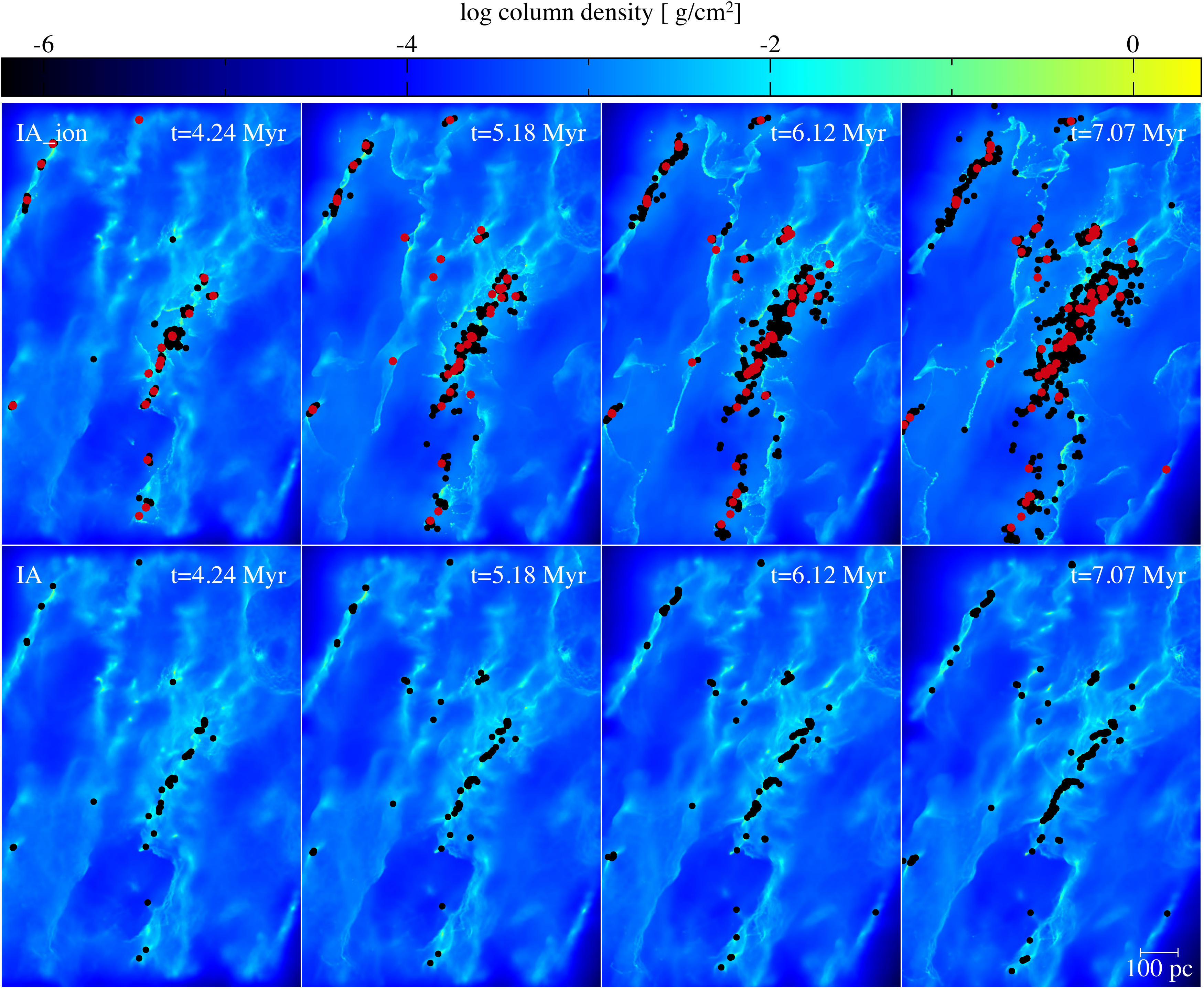}
  \caption{The column density is shown for the inter-arm region of the photoionisation run (top) and the no feedback run (bottom) over time. The frames start at 4.24 Myr when the first ionisation fronts are still fewer than 100 pc from their sources. Again, the ionisation appears to produce more sink particles, and sharper features in the gas, but this happens over longer timescales compared to the section of spiral arm. The larger red dots are ionising sink particles and the smaller black dots are non-ionising sink particles.}
  \label{fig:inter-arm_evol}
\end{figure*}

\begin{figure}
  \includegraphics[width=\columnwidth]{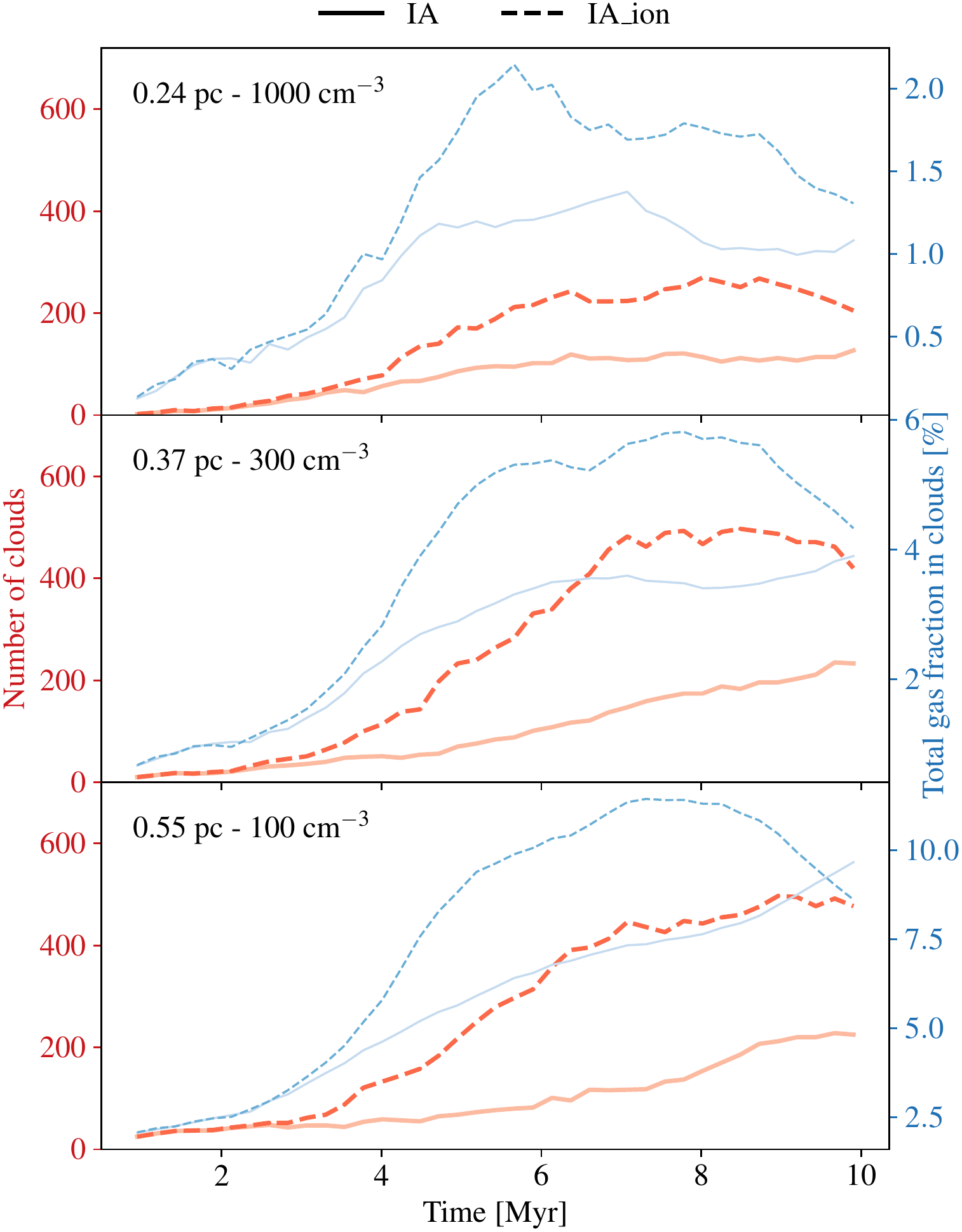}
  \caption{The evolution of the number of clouds (thicker red) and the gas fraction contained in them (thinner blue) is shown for the inter-arm region. This is equivalent to Figure \ref{fig:cloud_evol} for the spiral arm region. The evolution of the gas and clouds is similar to the spiral arm case, but occur over a longer timescale.}
  \label{fig:cloud_evol_inter-arm}
\end{figure}

The increase in SFR that ionisation produces in our simulations suggests a significant amount of triggered star formation in the surrounding interstellar gas. The triggering does not last indefinitely
in the simulations, presumably as dense gas on the verge of forming stars has already produced star formation, and further ionisation does not lead to any more collapse, and thus the star formation rate decreases. \citet{Dale2012} use the terms `weak triggering' and `strong triggering' to refer to temporary and permanent changes in the star formation process. They consider changes to the SFR weak triggering and changes in the SFE and number of stars (IMF) strong triggering. Since we find that photoionisation causes weak triggering that increases the SFR in all our runs, we refer to it as accelerated star formation throughout this paper.

As shown in Figure \ref{fig:sfr}, many more, and less massive, sinks form in the SR\_ion run. We identify three reasons for this:
\begin{enumerate}
\item Photoionisation clears away gas from the sinks shutting down accretion.
\item Groups of sink particles in feedback runs are less bound than with feedback (Section \ref{sec:clusters}); as a result of this and the relaxed sink particle merger criterion discussed in Section \ref{se:sinksandsampling}, there are fewer mergers in the feedback case.
\item in the SR\_ion run many additional sink particles form around ionisation fronts due to triggered star formation.
\end{enumerate}

In Figure \ref{fig:sinkdist} we estimate the distances over which this triggered star formation appears to occur in our simulations. For each sink particle in the runs with photoionisation we find the nearest sink at the equivalent timeframe for the no feedback run, and record these minimum separations. In the top panel in Figure \ref{fig:sinkdist} we plot the fraction of sink particles (by mass) in the SR\_ion model which lie above a given distance from any sink particle in the no feedback simulation. We see that in the SR\_ion run a significant mass of sink particles is situated more than 5 pc from the location of a sink particle in the SR run. It is important to note that we are not comparing where specific gas particles form sink particles or are accreted in runs with and without feedback, but rather the relative instantaneous locations of sink particles. As a result we do not identify displaced star formation that occurs near other sites of star formation, so this is an underestimate of the degree of triggering by photoionisation. There also may be sink formation at slightly different times in the two runs, which may overestimate the displacement on short timescales, but this should average out over time.

At its peak, between 3 and 5 Myr, the SFR in the SR\_ion run is 1.3 times that in the SR run; at this time the fraction of sink mass in the SR\_ion run greater than 5 pc from a sink in the SR run, is between 0.2 and 0.3 (as seen in Figure \ref{fig:sinkdist}). We infer from this that the majority of all accelerated star formation is occurring more than 5 pc from star formation in the no feedback case. In other words, most of the accelerated star formation in these simulations is occurring on scales larger than the majority of single cloud studies. 

The bottom panel of Figure \ref{fig:sinkdist} compares the 15 pc line from the top panel with the equivalent analysis in the other photoionisation runs. We see that the fraction of star formation triggered at distance is dependent on the total photoionising flux. Simulations with more flux produce more stellar mass further away from the equivalent no feedback run. It is worth noting that the SR\_los run only has a tiny fraction of stars at larger distances relative to the SR\_ion run. Additional triggering induced by long range photons at significant distances from star formation in the SR run appears to be present but only makes up a tiny fraction of the total star formation.

The formation of stars in the dense gas surrounding IFs - the collect and collapse model - has been explored analytically by \citet{Whitworth1994} and \citet{Hosokawa2006}. Simulators have looked at fragmentation in the IF for both uniform \citep{Dale2007} and fractal \citep{Walch2011} initial conditions. Observational evidence for the collect and collapse model is discussed in Section \ref{se:intro}. We make a distinction between collect and collapse triggering and the star formation triggered when ionisation fronts collide with dense gas structures or another ionisation front. There are multiple examples in the simulation of objects similar in appearance to, if much larger than, bright rimmed clouds which are created by radiation-driven implosion \citep{Bertoldi1989,Morgan2008}. We see a moderate amount of triggered star formation in such clouds when they are hit by shocks on one side. However, when clouds are compressed by HII regions on multiple sides, we see a much more significant acceleration in star formation.

Some comparable compression effects on clouds, in this case due to stellar winds and supernovae, have been noticed by \citet{Krause2018}. They present a `surround and squash' scenario in which they find that supernova feedback floods into low density regions created by earlier feedback and compresses the denser regions that the feedback flows around. This `sqaushing' effect induces further star formation.

The density maps in Figure \ref{fig:contour} show the last timeframe (3.3 Myr) at which the SFRs are comparable between the SR and SR\_ion runs (upper panels) and the time at which the SFR for the SR\_ion run peaks (4.24 Myr, lower panels). The ionisation runs on the right are overlaid with contours of the ionisation fraction. The SR run does eventually form stars in some of these regions, in one case the onset of star formation is delayed by $\approx$ 2 Myr and, once started, the process takes an additional $\approx$ 2 Myr. By comparison, this region is converted into stars in 1 Myr in the SR\_ion simulation, meaning that without photoionisation this star formation is delayed by 3 Myr. 

In Figure~\ref{fig:kenschmidt} we compare the star formation rates in our models with observations. We overplot the star formation rate versus surface density at different times for the different models on observations of star formation using the data from \citet{Bigiel2008}. In their work, \citet{Bigiel2008} measure surface densities and star formation rates in regions of 750 pc size scale, similar to our models. The SFR surface densities ($\Sigma_{\mathrm{SFR}}$) change as the total gas surface density decreases over time, an effect more pronounced in the SR\_ion run than the SR run. The combination data in Figure \ref{fig:kenschmidt} represents a weighted addition of the SR\_ion and IA\_ion simulations such that the total gas surface density ($\Sigma_{\mathrm{gas}}$) is equal to that for the whole galaxy from which our initial conditions are extracted.

The gas in our simulations is forming stars faster and more efficiently compared to nearby galaxies, as is common in most hydrodynamical simulations of star formation. This is also evident from the high absolute SFEs of over 30\%, shown in Figure \ref{fig:sfr}. These effects are exaggerated by the assumption that all mass in sink particles is stellar mass, although the SR\_50\% simulation goes a long way to addressing this issue.  Photoionisation only reduces the SFE on timescales longer than lifetimes of the majority of O stars. 

There are a number of reasons why our simulations overestimate the star formation rate. Firstly we only model one form of feedback, whilst ignoring supernovae, winds and radiation pressure. Around 3-15\% of sink particles are ionising, depending on the elapsed time and run, therefore, it is highly likely that accretion onto these non-ionising sinks would be significantly reduced by accounting for winds and/or radiation pressure. Supernovae are also likely to further reduce SFEs within the timescales of these simulations. A second, and probably more important, issue is that these simulations begin in the absence of any photoionising feedback. In reality there would be a population of ionising sources at the start of the simulation, which may diminish the extremes we see in the star formation rate. 

Furthermore our initial conditions contain no turbulence below the resolution of the galaxy simulations they were extracted from, meaning that the smallest size scale of turbulence in the initial clumps is of order 3-10 pc. Initial turbulence on these scales could lead to more fragmentation of dense gas structures early on and therefore lower SFRs and/or more sites of star formation. The accretion radius is also a factor in the high SFE, however, the SR\_2\ and SR\_2\_ion runs suggest that this is only a small effect.

\subsection{Cloud structure and morphology}
\label{sec:clouds}

In this section we look more quantitatively at the structure of the gas. We apply a friends of friends algorithm to identify clouds and clumps within the simulations. The algorithm groups particles within a certain distance of each other, and identifies them as a coherent structure. We also require that each structure has a minimum number of particles which we set to 100.  For the results here, we tested three different minimum distances, 0.24, 0.37 and 0.55 pc which lead to structures with slightly different ranges of masses and densities. Using a distance of 0.55 pc, we group particles into regions which are similar to molecular clouds, with median number densities of 100 cm$^{-3}$. For a distance of 0.24 pc, the structures picked out tend to be more similar to clumps within a cloud, and have median number densities of around 1000 cm$^{-3}$.

In Figure \ref{fig:massfn} we show mass functions for the clouds and clumps selected with number densities of 1000 cm$^{-3}$ distance criteria (top), 300 cm$^{-3}$ (middle) and 100 cm$^{-3}$ (bottom) for the simulations with and without ionisation. The mass functions are shown at a time of 4.24 Myr. From Figure \ref{fig:massfn} we see that there are more clouds and clumps in the simulation with photoionisation. Particularly there are more low mass clumps, but there are also a few more high mass clouds. Thus the  global effect of photoionisation appears to break up the clouds and clumps into smaller structures, however, in some cases the photoionisation causes the merger of two or more clouds (i.e. the largest cloud in middle panel). We can see this a little in Figures \ref{fig:wholesim} and \ref{fig:ionregion}, where the ionisation tends to make the filaments thinner and denser, and in some cases break up the filaments. 

We perform the two-sample K-S test on the cloud distributions for the SR\_ion and SR\_50\% runs, each relative to the SR run but at multiple epochs. We find that at 4.24 Myr (Figure \ref{fig:massfn}) the p-values from the K-S test for the SR\_ion run are 0.008, 0.00003 and 0.34 for clouds of median number density 1000, 300, and 100 cm$^{-3}$ respectively. This suggests that clouds as defined by our middle criterion, median number density 300 cm$^{-3}$, are the most disrupted and clouds as defined by a median number density of 100 cm$^{-3}$ are less affected. This trend is supported consistently throughout the time evaluation of the simulations. The objects defined by the densest criterion (clumps) are affected less than those by the middle criterion; this could be due either to dense gas resisting feedback or because these objects are dominated by the process of sink particle formation which does not vary between simulations. The K-S test results for the SR\_50\% run suggest a similar trend -- as expected, the significance of the cloud disruption is lower in this run.

We show the global cloud properties over time in Figure \ref{fig:cloud_evol} for the same simulations and cloud definitions as in Figure \ref{fig:massfn}. The peak difference in cloud numbers between the no feedback and photoionisation runs for all cloud densities is around a factor of 2. This difference is not produced solely by breaking up of clouds but also by the formation of new clouds under pressure from HII regions. Figure \ref{fig:cloud_evol} also shows that photoionisation causes additional gas to be agglomerated into clouds, both newly formed and pre-existing. This effect is far more pronounced in the densest objects. At the peak in both the SR\_ion and SR\_50\% runs there is approximately twice as much gas in clumps (clouds of number density 1000 cm$^{-3}$). In the SR\_ion runs the total gas fraction peaks at around 3.3 Myr, the peak in the number of clouds is at around 4.2 My. The SR\_50\% run shows very similar behaviour but it is delayed by around 1 Myr. 

In Figure \ref{fig:clump_map} we overplot the clumps found with the 0.55 pc distance criteria in the SR, SR\_ion and SR\_los runs. In the no feedback model, there is clearly one very massive cloud, whereas, with ionisation, this cloud has been broken up into multiple structures. Furthermore for the cases with ionisation - particularly the SR\_los run -  we see clumps in new regions often in spurs off the spiral arm. Again this is likely due to the impact of photoionisation compressing these regions. We find minimal difference between the statistical properties of the SR\_ion and SR\_los runs and conclude that for an isolated section of spiral arm, long range ionisation is not critical. We use line of sight limits that are as short as possible, while still being comfortably larger than the typical initial separations of clumps and clouds for each set of initial conditions. \citet{Emerick2018} examined the importance of long-range ionisation in low-mass dwarf galaxies, they conclude that long-range ionisation is particularly important in driving galactic winds.

Global galaxy simulations indicate that stellar feedback is an important source of kinetic energy in the ISM and necessary for preventing GMCs from becoming too gravitationally dominated \citep{Dobbs2011}. In Figure \ref{fig:virial} we investigate whether the ionisation changes the ratio of the kinetic to gravitational energy (i.e. virial parameter $\alpha$) in the clumps and clouds in our simulations. We show results for the distance criteria of 0.24 pc (top) i.e. smaller clumps, 0.37 pc (middle) and 0.55 pc (lower), which includes larger molecular clouds. As well as the SR and SR\_ion simulations, we also include the 50\% ionisation model, SR\_50\%. In all cases, we see that ionisation has a notable effect on the virial parameter of the clouds. The distribution of $\alpha$ shifts to higher values; the clouds and clumps with these high values may well be very transient. For the clumps (top panel Figure \ref{fig:virial}), there are still many bound objects in the cases with ionisation. However, for the clouds (bottom panel), the numbers of bound clouds are noticeably less with ionisation, and overall the clouds tend to be unbound. We find this is still true for the SR\_50\% simulation. Even though there is a smaller and perhaps more realistic amount of ionisation than is modelled in our fiducial case, this still has a significant impact on the gas. 

\subsection{Stellar clusters}
\label{sec:clusters}

Whilst we mostly focus on the gas distribution, especially since our sink particles are fairly massive so that clusters are not resolved with a large number of sink particles, we nevertheless provide a short analysis of the effects of photoionisation on stellar clusters in the simulations. As for our analysis of the cloud structure, we focus on the SR, SR\_ion and SR\_50\% simulations. We make no attempt to identify clusters based on the origins of the sink particles. Instead we build a minimum spanning tree of all of our sink particles and cut all branches of this tree above some distance. We call this the cut-off distance. Repeating this for a range of cut-off distances allows us to assess multiple sets of clusters defined by different size scales for the same set of sinks. We show in Figure \ref{fig:mass_bound_clusters} cluster properties versus this cut-off distance, which is a proxy for the size scale of the clusters.

In Figure \ref{fig:mass_bound_clusters} (top panel), we show the average mass of the cluster versus the distance criterion. The figure indicates that clusters in the model with no feedback (SR) have higher masses, and therefore densities, than their counterparts in the simulations with ionisation. In the lower panel of Figure \ref{fig:mass_bound_clusters} we show the fraction of bound clusters for the different cut-off distances. As would be expected with short cut-off distances the clusters tend to be bound in all the models, although there are fewer bound clusters with ionisation. As the cut-off distance increases, the fraction of bound clusters stays fairly constant with no feedback, but drops off by around 20\% in the models with ionisation. So, similarly to the behaviour for the gas, the ionising feedback appears to lead to fewer bound clusters. And again, the model with 50\% ionisation, SR\_50\%, shows a significant difference to the no feedback case, indicating that smaller amounts of ionising feedback can still have a significant impact.

Observations suggest that 40 to 90 \% of stars form in clusters, depending on the criteria for a cluster \citep{Bressert2010}, although \citet{Kruijssen2012} suggest the fraction is less than this, around 30\%. The latter estimate would mean that we form too many bound clusters, although we again note that the stellar accretion and mergers are not well resolved or represented in our simulations, and again we may be missing other forms of feedback which may influence cluster evolution. While our bound fractions are unlikely to be accurate, the tendency towards a significantly lower fraction of bound clusters in the presence of photoionisation is very clear. 

\subsection{Comparisons with an inter-arm region}
\label{sec:inter-arm}

The inter-arm region evolves very differently in the presence of photoionising radiation, as shown in Figure \ref{fig:inter-arm_evol}. We again see the wider distribution of star forming regions, and, in this case, we can clearly see photoionisation causing a more filamentary structure in the SR\_ion run.

In our spiral arm region (SR), all of the runs show that photoionisation accelerates star formation in the surrounding gas. Since a key difference between the IA and SR initial conditions is the separation of the star forming regions, it is interesting to explore this effect in these regions. In Figure \ref{fig:cloud_evol_inter-arm} we plot the  number of clouds, and total fraction of gas in clouds for the inter-arm region, as previously presented in Figure \ref{fig:cloud_evol} for the spiral arm region. We see increases in the total gas in clouds, number of clouds and star formation rates when photoionisation is included (IA\_ion), however, not in the same proportions as in the spiral arm. In both the simulations with and without feedback (IA and IA\_ion runs), the SFE has not begun to plateau after 10 Myr, whereas the SR\_ion run has reached 90\% of its final SFE by 7 Myr.

\citet{Duarte-Cabral2016} explore the differences between arm and inter-arm clouds in very similar simulations to these, but lacking photoionisation, by analysing the simulations presented in \citet{Dobbs2015}. They find that while predominantly clouds are very similar in the arms and inter-arms, there are some variations. Of interest to this work, they find that the most filamentary clouds are found in the inter-arms and suggest gravitational shear as the main cause of this.

We see the same effect in our simulations, but notice how photoionisation further defines the shape of these filaments and increases their mean density. By comparison of Figures \ref{fig:cloud_evol} and \ref{fig:inter-arm_evol}, we see a greater total increase in gas in clouds with feedback in the inter-arm runs relative to the arm runs, this additional dense gas is also longer lasting. The additional definition to the filaments seen in Figure \ref{fig:inter-arm_evol} is unmistakable.

We compare velocity dispersions of GMCs between the IA, IA\_ion, SR, and SR\_ion runs at the times at which the total gas in clouds peaks for the runs with photoionisation (3.3 Myr for SR and 5.9 Myr for IA). We find that GMCs (clouds of median number density 1000 cm$^{-3}$) have higher velocity dispersions in the arms than inter-arms which, like \citet{Duarte-Cabral2016}, is in agreement with observations of M51 by \citet{Colombo2014} and within the Milky Way by \citet{Rigby2019}; this difference is reduced in the feedback runs. We notice similar behaviour in clumps although it is less pronounced and reduced less by photoionisation.

While the increase in total gas in clouds in the IA\_ion run is greater than the SR\_ion run and the number of clumps is increased by a similar amount, the effect on the SFR is much less stark. The peak in the SFR in the feedback case being only 1.3 times higher than without feedback. The SFR per unit area varies between 10 and 20 times smaller in the inter-arm region compared to the spiral arm over time.

Simulating an inter-arm region with supernovae in addition to photoionisation will be particularly interesting owing to the longer time-scale of gas depletion and limited dense gas to absorb the feedback.

\section{Conclusions}

We have presented comparisons between simulations, with and without photoionising feedback, of 0.5 kpc$^2$ regions extracted from galaxy scale simulations. Our simulations contain a number of molecular clouds, and ionising feedback is able to affect the surrounding gas and neighbouring clouds. We are also able to study the behaviour of ionisation in a region with many clouds, i.e. the spiral arm, versus a relatively low density inter-arm region. We focus predominantly on three cases: no feedback, photoionising feedback and a prescription with half the total Lyman flux. We find that the interactions between ionisation fronts and neighbouring clouds are highly influential on the star formation rate, cloud distribution and cloud dynamics, but only minimally affect the overall star formation efficiency. In particular we find that:

\begin{enumerate}
  
\item Star formation is accelerated by photoionising radiation largely due to the pressure of HII regions on clumps/clouds that would not otherwise form stars for some time, often at large distances from sites of star formation in runs without feedback. This process occurs over timescales of several Myr. After this period of acceleration, star formation decreases rapidly because much of the dense gas is used up. Thus we find that ionisation produces a significant amount of triggered star formation in our simulations. 
\item The number of cluster sink particles is nearly doubled by the effects of photoionisation, but the accretion on to them is much less than in the no feedback case. The latter seems in basic agreement with simulations by \citet{Gatto2017} who find that winds reduce accretion on to clusters and limit their mass and \citet{Zamora-Aviles2019} who find the same for photoionisation.
\item Long range ionisation causes the formation of clouds at very high distances (hundreds of pc) from the nearest ionising sink particle. However, these clouds contain only a tiny proportion of the total star formation. When limiting the maximum distance over which ionisation can act, we see that reducing the range of ionising photons to 100 pc does not modify the global statistical properties of gas and star formation in the spiral arm.  
\item Photoionisation increases both the number of clouds and clumps and to a lesser extent the total gas fraction within clouds. This increased fraction of gas in clouds indicates that fragmentation is not the only cause of the increased number of clouds, but also that new clouds are being created, or at least having their collapse accelerated, by pressure that results from photoionisation.
\item Including photoionising feedback produces stellar clusters which are less compact, and heavily reduces the fraction of bound clusters relative to the no feedback case.
\item In inter-arm regions, cloud evolution and star formation follow a similar pattern to the spiral arm region but on a larger scale and with a smaller increase in the star formation rate. Although the distances between star forming clumps are larger in the inter-arm regions, this is offset by the lower densities between clouds, allowing ionisation fronts to travel large distances.
\item Final star formation efficiencies are only modestly affected by photoionisation, however, the conversion of gas into stars occurs in around half the time in the runs with photoionisation.
  
\end{enumerate}

Compared to previous work, the ionisation has a greater effect of enhancing star formation in our simulations, although in common with previous simulations, we also see ionisation breaking up and disrupting the clouds. One of the main reasons for the increased star formation is that our clouds are not isolated, and unlike previous work \citep{Dale2014,Geen2016,Ali2019,Zamora-Aviles2019}, the feedback does not escape into an empty region, but impacts other material. We do see a large decrease in star formation, but only after a few Myr due to much of the dense gas being used up. 

We may see the gas replenished, as further gas flows into the spiral arms, and star formation increasing again. Potentially, as spurs or clouds of gas enter the spiral arms, this could produce a cyclic pattern of star formation in spiral arms with gas undergoing more rapid star formation and depletion followed by a quiescent period.

Although our simulations predict an increase in star formation with ionisation, there are caveats to this result. Firstly, we do not include any other forms of feedback, which may lead to a decrease in star formation earlier, or in the case of supernovae, be more destructive than ionisation. Furthermore, in reality, there would be a population of ionising stars at the outset of our simulation which are affecting the gas. Inclusion of these is likely to avoid the large peak in star formation seen here, and again lead to a decrease in the star formation earlier in the simulations. Another possible improvement would be to include turbulence on the smallest scales when setting up our initial conditions.

The role that magnetic fields would play in this work is far from certain. In general magnetic fields are thought to have little impact on the early development of HII regions, however, they can certainly distort the shape of HII regions as they grow \citep{Krumholz2007,Mackey2011}. \citet{Geen2015a} run hydrodynamic simulations of clouds with both photoionisation and a magnetic field. They find that the magnetic field does not strongly affect the global properties of the HII region although the structure is noticeably affected. By inference from this limited body of work we suggest that on these kpc scales magnetic fields are unlikely to dominate the effects of photoionisation, but are certainly an important part of the complete picture.

\section*{Acknowledgements}

The authors would like to acknowledge the use of the University of Exeter High-Performance Computing (HPC) facility in carrying out the no feedback simulations. The photoionisation simulations were performed using the DiRAC Data Intensive service at Leicester, operated by the University of Leicester IT Services, which forms part of the STFC DiRAC HPC Facility (www.dirac.ac.uk). The equipment was funded by BEIS capital funding via STFC capital grants ST/K000373/1 and ST/R002363/1 and STFC DiRAC Operations grant ST/R001014/1. DiRAC is part of the National e-Infrastructure. CLD acknowledges funding from the European Research Council for the Horizon 2020 ERC consolidator grant project ICYBOB, grant number 818940. MRB acknowledges support from the European Research Council under the European Community's Seventh Framework Programme (FP7/2007-2013 Grant Agreement No. 339248). The column density figures were made using SPLASH \citep{Price2007}. 

%%%%%%%%%%%%%%%%%%%%%%%%%%%%%%%%%%%%%%%%%%%%%%%%%%



\bibliographystyle{mnras}
\bibliography{library,textbooks}

%begin{thebibliography} {Sternberg2003}
%\bibliographystyle{mnras}
%\bibliography{library.bib}
%\end{thebibliography}

%%%%%%%%%%%%%%%%%%%%%%%%%%%%%%%%%%%%%%%%%%%%%%%%%%

%%%%%%%%%%%%%%%%% APPENDICES %%%%%%%%%%%%%%%%%%%%%

\appendix

\section{Increasing particle resolution}
\label{app:part_resolution}

Here we describe our method for increasing the resolution in smoothed particle hydrodynamics (SPH) for the purposes of re-simulation (rather than modifying resolution locally during a simulation). All the methods which we discuss here, along with our own, involve distributing $N$-1 new daughter particles around each original parent particle and reducing the mass of all particles by a factor $N$; daughter particles inherit all other quantities from their parent. The methods vary in the number and locations of daughter particles.
 
\citet{Dobbs2015} overlay a body centred cubic lattice with the original particles in the centre. The cubes are of side length 1.2$h$, where $h$ is the SPH smoothing length of the original particle, which leads to the daughter particles sitting $0.6\sqrt 2 h$ from the original particle which is just under half way across the region of compact support. This process is repeated a second time giving a final resolution increase factor of 81. 

\citet{Kitsionas2002} use a similar approach, but for increasing resolution on the fly locally; they use a hexagonal close-packed grid consisting of 12 daughter particles. They place these particles 1.5$N^{-1/3}h$ from the parent particle which places these particles slightly closer to the parent than in \citet{Dobbs2015}.

Both these methods leave a clear visual imprint on the initial conditions, however, after a short relaxation time the density profile is smoothed. They have the distinct advantage that they are incredibly simple to implement, however, they are not so well suited for large increases in resolution (we use $N=311$ and 823). For large $N$ they require multiple iterations and provide a limited set of possible values for $N$. 

\citet{Rey-Raposo2014} distribute $N$-1 particles within $2h$ of each original particle. The daughter particles are placed through an inverse sampling method. They approximate the truncated SPH kernel to a Gaussian since its inverse function is simpler to work with. This means that the radial positions of their added particles follow the SPH smoothing kernel's density distribution. This has the advantage of working for any value of $N$. While their sampling method avoids large clumps of particles there is still a stochastic element to the locations of the daughter particles which we find exaggerates the extremes in density immediately after the resolution increase, although again this is largely smoothed in a short relaxation time. 

Since this work is interested in star formation we are keen to avoid over-densities in the initial conditions since they may lead to too many star forming clumps forming in the early stages of the simulation. We have developed a method that uses a spherical grid that does not leave an imprint and almost entirely eliminates over-densities.

We build our grid in $q$-space, where $q$ is a dimensionless distance linked to physical space by
\begin{equation}
q = \frac{r}{h},
\end{equation}
where $r$ is the physical distance from the centre of the grid (the site of the original particle). Defining truncated kernel functions in terms of $q$ is common in the SPH community \citep{Monaghan1992,Price2012}. This means that we can use the same grid for all parent values of $h$. The radius of compact support for the SPH smoothing kernel (in this case the M4 cubic spline) is 2$h$ which translates to 2$q$ in $q$-space, this is the maximum allowed size of the grid.

\begin{figure}
  \centering
  \includegraphics[width=\columnwidth]{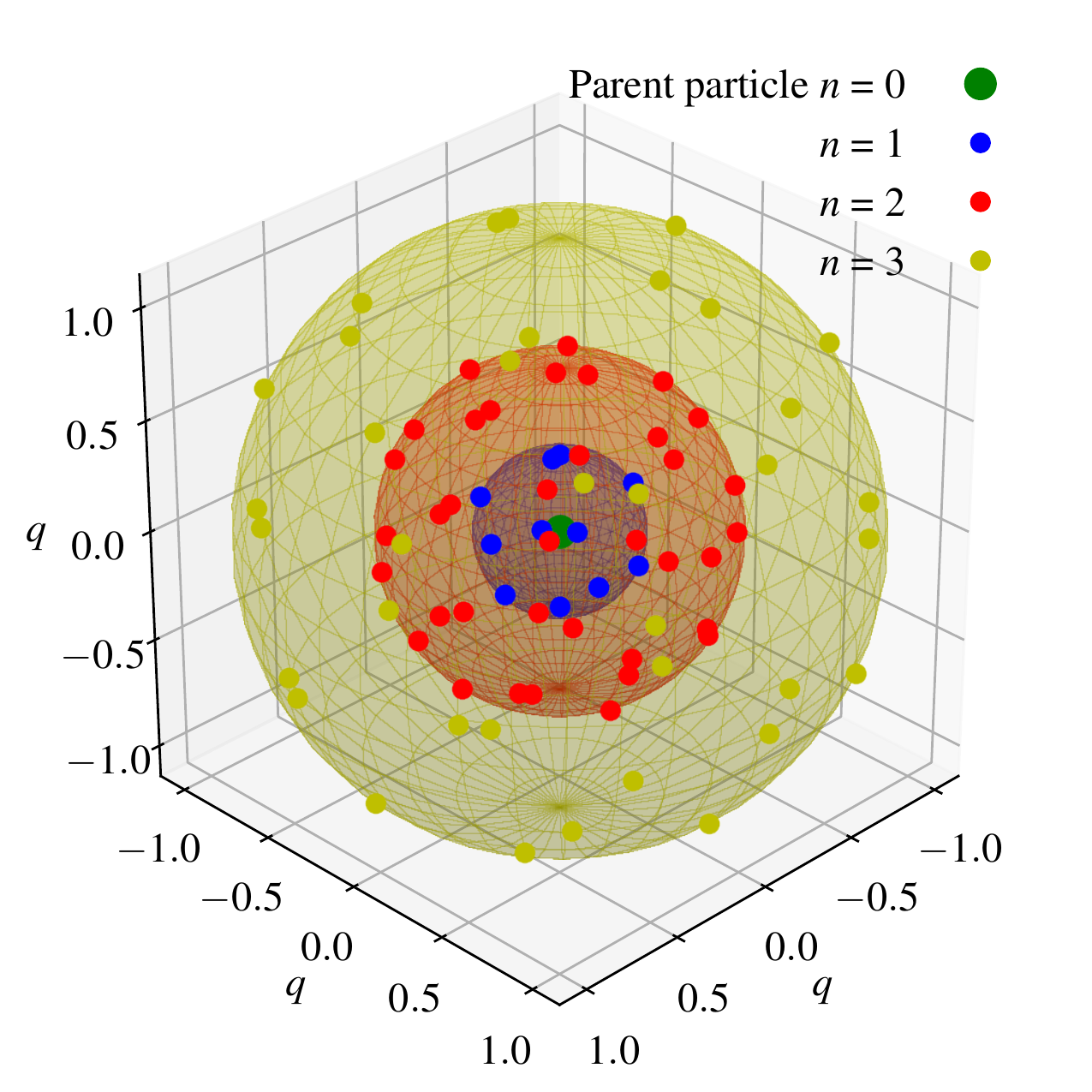}
  \caption{A 3-D visualisation of the concentric shell structure of the resolution increase grid. For this plot we have chosen $N=85$, rather than the 311 used in most of our simulations, for clarity. The shells $n$ = 1-3 hold 11, 37 and 36 particles respectively and are situated at radii of 0.35$q$,0.73$q$ and 1.30$q$.}
  \label{fig:spheres}
\end{figure}

The grid is made up of particles sitting on a set of concentric shells, each assigned a shell number $n$. The grid requires an axially uniform distribution but a radial dependence that causes the effective density to drop off with the smoothing kernel. An example of one of our grids is plotted in Figure \ref{fig:spheres}, in this example $N=85$. To characterise a grid we need to define two quantities, the radius of each shell, $q_n$, and the the nearest neighbour separation on each shell, $s_n$. The site of the original particle is denoted by $n=0$, therefore $q_0$ is equal to zero, however, $s_0$ can have any value greater than 0 and less than 2$q$. 
%We discuss the choice of $s_0$ and the placement of particles on a shell later in the section. 

The radial drop in effective density is achieved by calculating the nearest neighbour separation, $s(q)$, using the normalised inverse of the kernel function, $w(q)$, as the fractional increase above $s_0$ with increasing radius 
\begin{equation}
  s(q) = s_0\left(\frac{w(q)}{w(0)}\right)^{-\frac{1}{3}},
  \label{eq:norm_inverse}
\end{equation}
where $w(0)$ is the smoothing kernel evaluated at $q_0$ -- the centre of the grid. 

\begin{figure}
  \centering
  \includegraphics[width=\columnwidth]{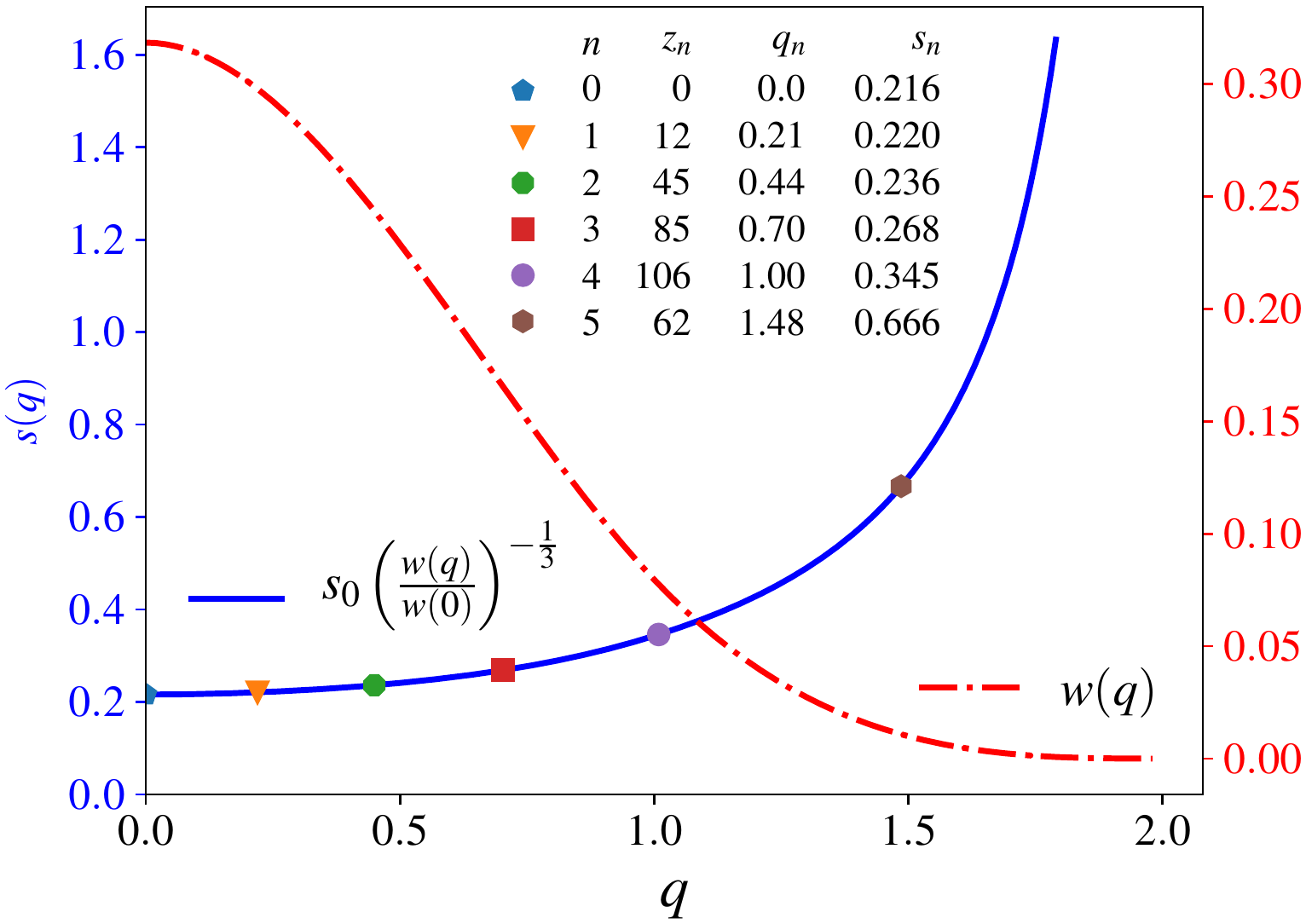}
  \caption{The M4 cubic spline (red dot-dash line and right hand axis) and its normalised inverse cubic (blue solid line and left hand axis). The right hand axis values are arbitrary since the value of the kernel is dependent on the smoothing length. The normalised curve is multiplied by the characteristic nearest neighbour separation ($s_0$) at the centre of the grid so that the left axis gives the separation at any given radius ($s_q$). In this example $s_0$ is 0.216$q$, this gives $N=311$. The legend table lists the following shell properties: shell number ($n$), number of particles ($z_n$), radius ($q_n$), and the nearest neighbour separation ($s_n$).}
  \label{fig:shell_radii}
\end{figure}

Particle separation is constant on each of the shells in the grid so the $s_n$ values can be easily calculated from \ref{eq:norm_inverse}. However, there is some choice in how we determine the separation of neighbouring shells. We define this separation as the mean value of $s(q)$ between those shells
\begin{equation}
  q_{n}-q_{n-1} = \frac{1}{q_{n}-q_{n-1}}\int^{q_{n}}_{q_{n-1}}s(q){\rm d}q.
  \label{eq:mean_sep}
\end{equation}
We solve Equation \ref{eq:mean_sep} by an iterative numerical process to find $q_{n}$. This is preferable since the computation time minimal and the analytic solution is non-trivial. We repeat this process for each subsequent shell until $q_{n}$ is greater than 2$q$, discarding the last result. Figure \ref{fig:shell_radii} shows $w(q)$ and its normalised cubic inverse (Equation \ref{eq:norm_inverse}). The shell locations $q_n$ are marked for $s_0=0.216q$ which gives $N=311$. 

The shells are populated by $z_n$ particles whilst ensuring that every particle is as far as possible from its nearest neighbours. We approximate the desirable nearest neighbour separation for a number of particles on a shell as
\begin{equation}
  2q_n\sin\sqrt\frac{\pi}{z_n},
  \label{eq:sep_approx}
\end{equation}
choosing $z_n$ to achieve the best match between $s_n$ and this value. This means that the choice of $s_0$ defines the number of particles in a grid. We solve iteratively for many $s_0$ values to create grids for new values of $N$.

There are many ways of distributing particles evenly over the surface of a sphere. \citet{Raskin2016} place particles on the vertices of platonic solids in an approach for building spherically symmetric initial conditions for SPH that follow a radial density dependence. This has the advantage of offering complete symmetry with every particle's situation on the shell being identical. However, this platonic solid approach does not allow for any number of particles to be placed on a shell. It is also possible to use spiral methods, such as a spherical Fibonacci point set, to place any number of particles on a shell. These spiral approaches lead to slight variations in each particle's relative environment on the shell. This variance depends on both $N$ and the specific spiral method. We use an `electron repulsion' approach in which we allow the particles on the shell to repel each other until they approach an energetic minimum.  

We avoid any imprinting of the grid on the new initial conditions by generating 72 grids and cycling through them for each particle. Randomly rotating the grid in spherical polar coordinates for each parent particle would achieve the same end. We make an estimate of the new smoothing lengths for each daughter particle and, finally, re-calculate the values of $h$ by interpolation. 

\section{Mass Resolution}
\label{app:mass_resolution}

\begin{figure}
  \includegraphics[width=\columnwidth]{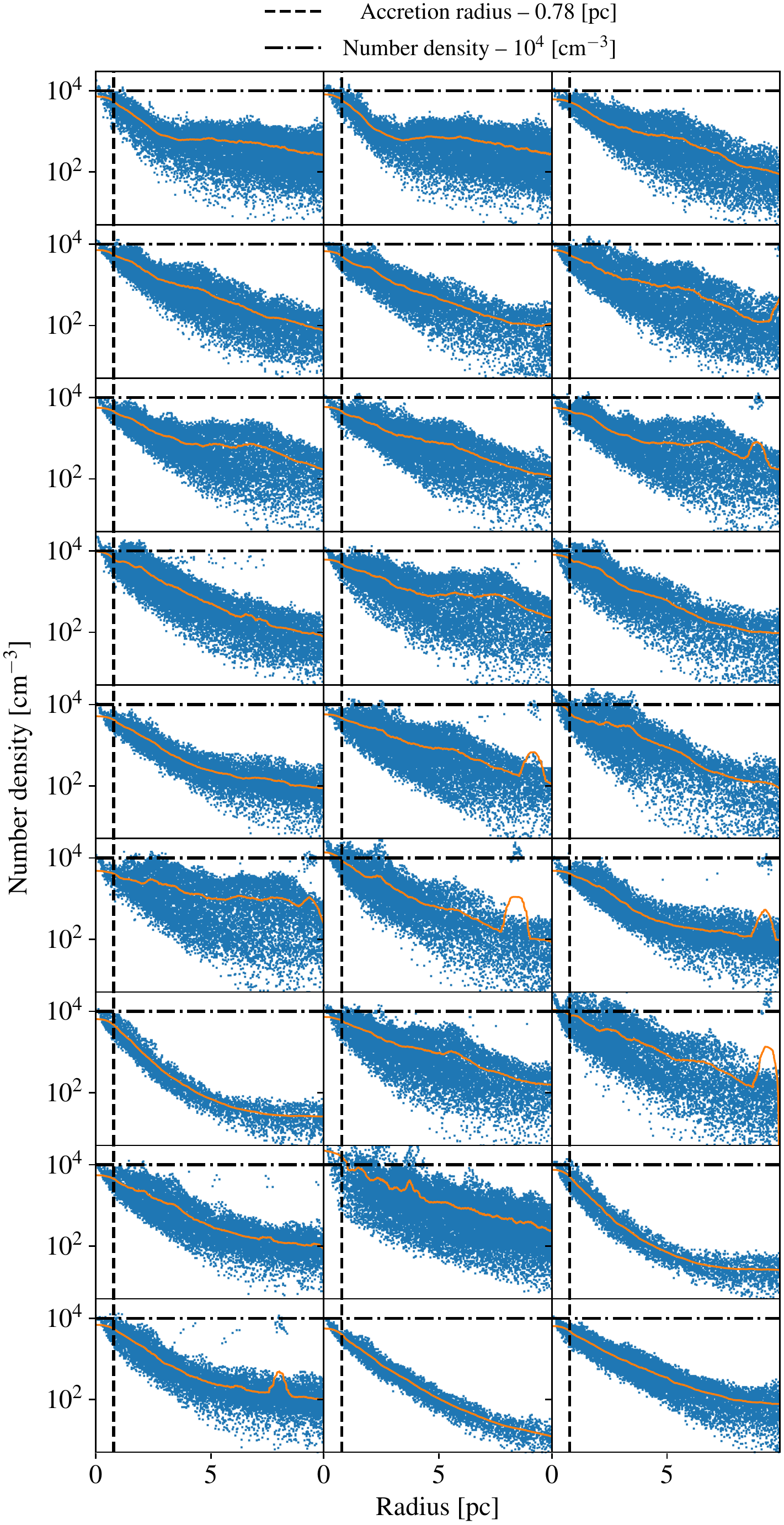}
  \caption{Here we show the surroundings of the first 27 cluster sink particles immediately before they form in the SR\_ion run. The blue dots each represent an SPH particle and the orange line is a mean of the radial profile. The vertical dashed lines mark the accretion radius (0.78 pc) of the sink particles and the horizontal dot-dash lines are at 10$^4$ cm$^{-3}$, which appears to be a characteristic particle number density upper boundary around nearly all of our sink particles at formation.}
  \label{fig:sinkprox}
\end{figure}

\begin{figure}
  \includegraphics[width=\columnwidth]{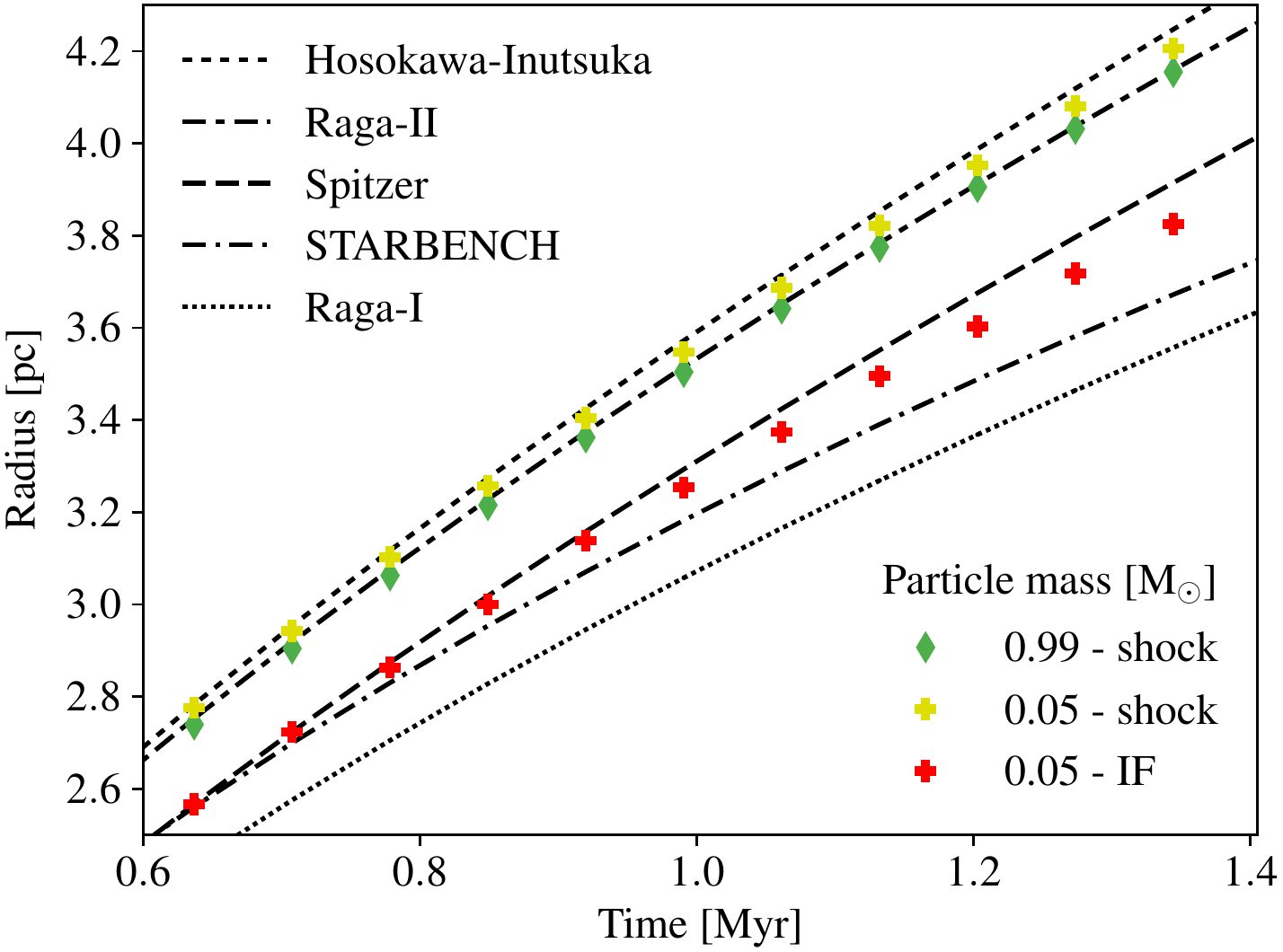}
   \par\bigskip
  \includegraphics[width=\columnwidth]{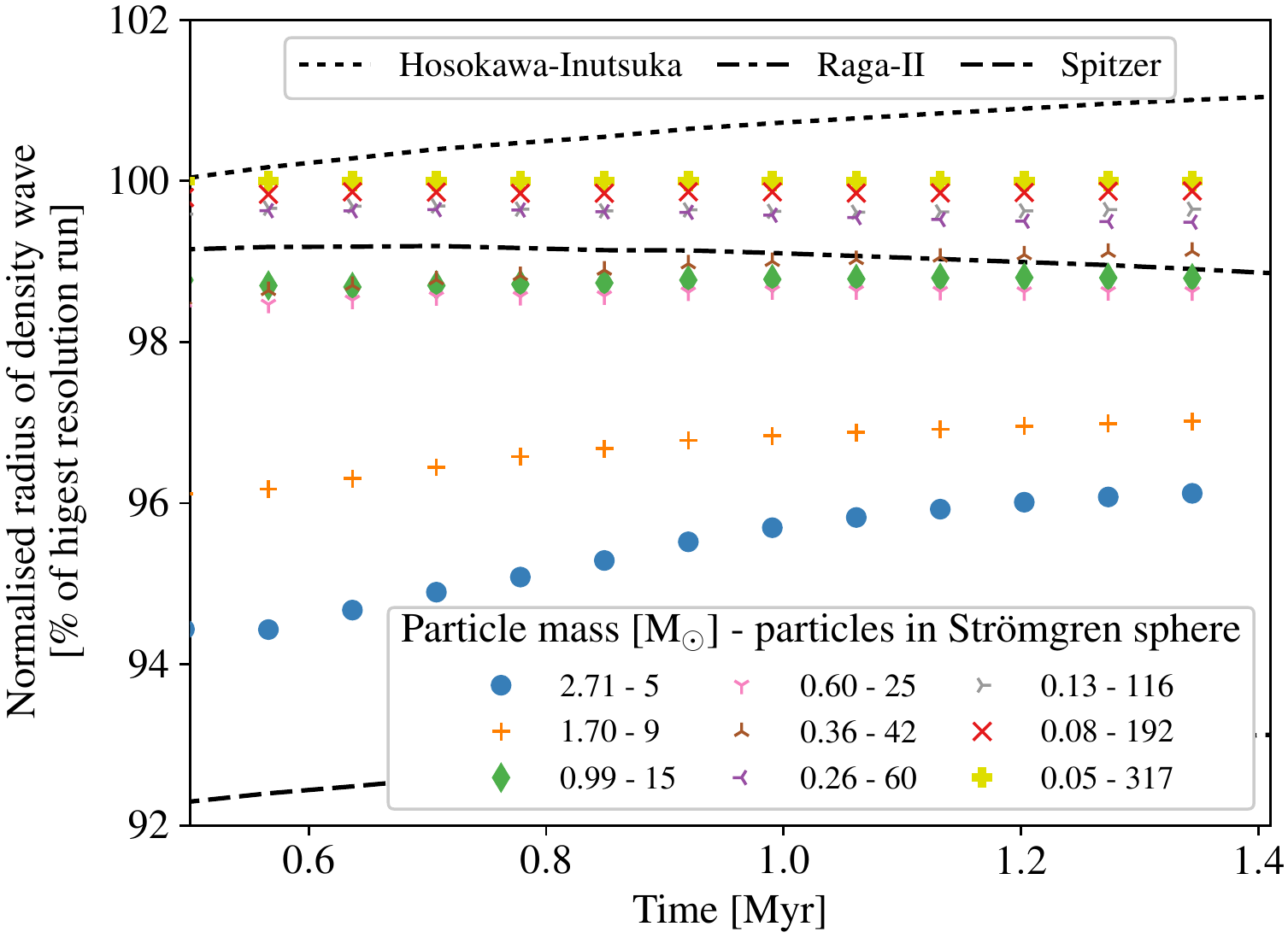}
  \caption{The top panel shows the D-type expansion of an HII region around a source of Lyman flux 5 $\times\ 10^{49}$s$^{-1}$ for simulations at two resolutions. The diamonds (dark green) at a resolution comparable to our fiducial runs ($\approx$ 1M$_\odot$ per sph particle) in this paper and the crosses (light green) at the highest resolution test run (20 times more particles) a number of analytical solutions are also plotted. We also plot the location of the ionisation front approximated using the ionised fractions of gas particles for the highest resolution run (red crosses). The lower panel shows the same shock data normalised using the 0.05 M$_\odot$ run, including 7 additional resolutions. Each run is defined by its mass per particle and the number of SPH particles that make up the initial Str{\"o}mgren sphere after the R-type expansion phase. The initial uniform number density in all these simulations is $10^4$ cm$^{-3}$.}
  \label{fig:dens}
\end{figure}

\begin{figure}
  \includegraphics[width=\columnwidth]{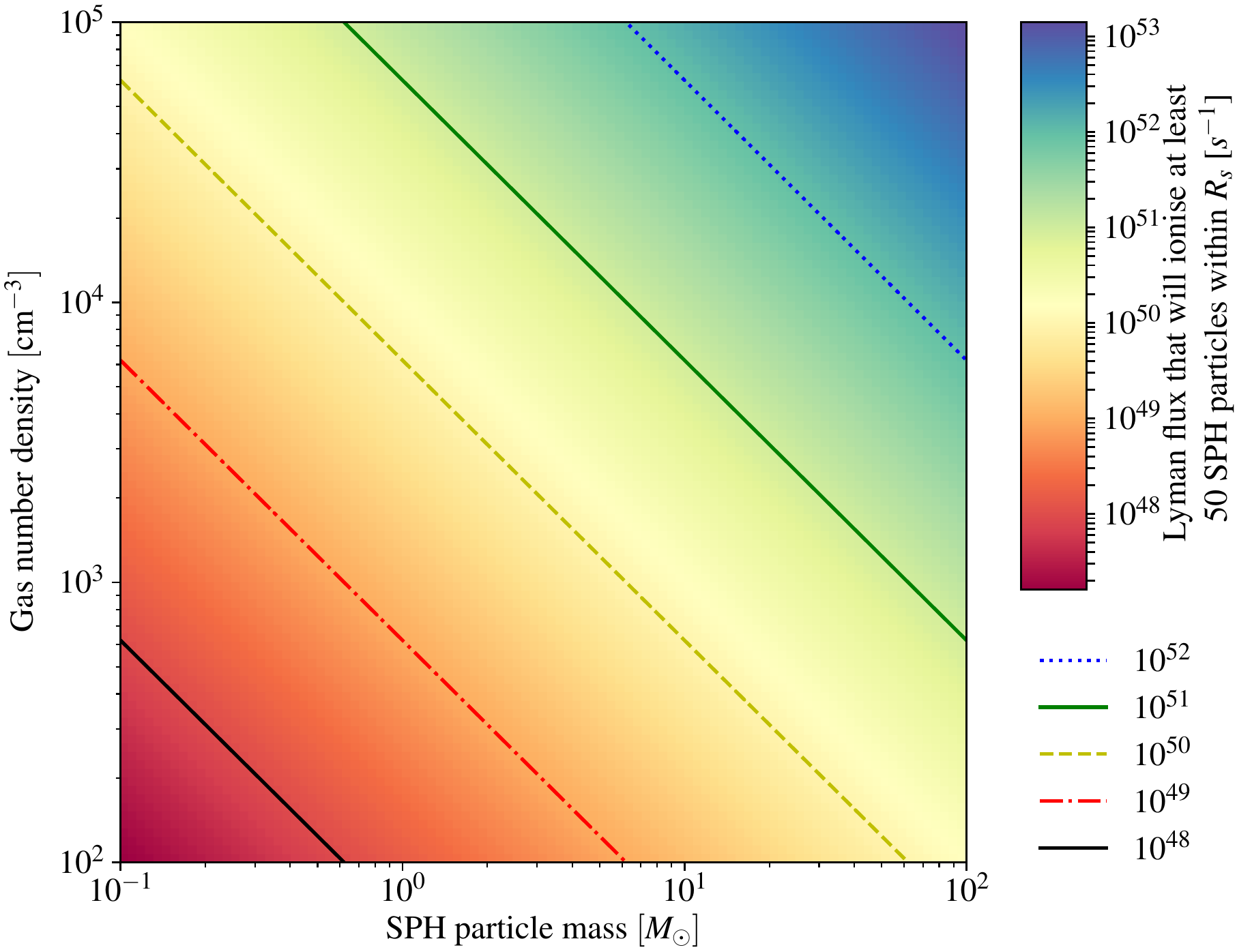}
  \caption{A plot indicating the locations at which Str{\"o}mgren spheres will contain mass equal to 50 SPH particles in gas density -- resolution space for a range of Lyman fluxes. The lines are contours for Lyman flux.}
  \label{fig:50sphion}
\end{figure}

The challenges of implementing a rigorous photoionisation algorithm at resolutions low enough to simulate such a large region are significant. We include this section to quantify the limitations and uncertainties that result from our choice of resolution. To do this we run a set of single source test simulations in uniform density with no velocity field at a variety of mass resolutions. We choose an initial number density of $10^4$ cm$^{-3}$, since this is the typical environment in which Str{\"o}mgren spheres form within our simulations, as shown in Figure \ref{fig:sinkprox}. We use an ambient temperate of 100 K which is slightly higher than our simulations but reduces the computational expense of these tests.

We found the STARBENCH paper \citep{Bisbas2015} to be very useful in testing our photoionisation method. We use their nomenclature to identify the following analytical solutions to the D-type expansion phase, they are each dependent on the Str{\"o}mgren radius ($R_{\rm S}$ Equation \ref{eq:rs}):

\begin{itemize}

\item Raga-I -- \citet{Raga2012a} consider a `thin' shock to be in pressure balance with the ionised gas inside the IF and derive the IF radius over time by considering the relative velocities of the shock and the material outside,
\begin{equation}
  \frac{1}{c_{\rm i}}\frac{{\rm d}R_{\rm RI}(t)}{{\rm d} t} = \left(\frac{R_{\rm S}}{R_{\rm RI}(t)}\right)^{3/4} - \frac{c^2_{\rm o}}{c^2_{\rm i}} \left(\frac{R_{\rm S}}{R_{\rm RI}(t)}\right)^{-3/4},
  \label{eq:Raga-I}
\end{equation}
\noindent
where $c_{\rm i}$ and $c_{\rm o}$ are the sound speeds inside and outside the IF and $t$ is time. It is important to note that fully ionised gas has half the mean molecular weight compared to completely neutral gas.

\item Spitzer -- The Spitzer solution \citep[p. 333]{Spitzer1978physical} can be reached by ignoring the right hand term in Equation \ref{eq:Raga-I} which is small at early times,
\begin{equation}
  R_{\rm Sp} = R_{\rm S}\left(1+\frac{7}{4}\frac{c_{\rm i}t}{R_{\rm S}}\right)^{4/7}.
\end{equation}

\item Raga-II -- \citet{Raga2012b} add the momentum of the expanding shell to their model from \citet{Raga2012a},
\begin{equation}
  \ddot R_{\rm RII}+\left(\frac{3}{R_{\rm RII}}\right)\dot R^2_{\rm RII} = \frac{3R^{3/2}_{\rm S}c^2_{\rm i}}{R^{5/2}_{\rm RII}}-\frac{3c^2_{\rm o}}{R_{\rm RII}}.
\end{equation}

\item  Hosokawa-Inutsuka -- \citet{Hosokawa2006} derive the radius from the equation of motion of the shock, but do not consider the pressure outside the IF,
\begin{equation}
  R_{\rm HI} = R_{\rm S}\left(1+\frac{7}{4}\sqrt{\frac{4}{3}}\frac{c_{\rm i}t}{R_{\rm S}}\right)^{4/7}.
\end{equation}

\item \citet{Bisbas2015} present a semi-empirical equation of their own referred to as the STARBENCH equation intended as a tool for benchmarking, for further detail we recommend referring to their paper.

\end{itemize}

The location of the ionisation front is difficult to accurately identify using the ionisation fraction of each SPH particle, particularly when the resolution is poor. Instead we approximate the radius of the resulting shock using a weighted average of the radii of all the particles (N) above the ambient density ($\rho_0$),

\begin{equation}
R_{\mathrm{shock}} = \frac{1}{\rm N}\sum_{i=1}^{\rm N}{(\rho_i-\rho_0)r_i},
\end{equation}

\noindent
where $\rho_i$ and $r_i$ are the density and radius of each particle in the shock above $\rho_0$ respectively. We plot these radii for a number of resolutions over time in Figure \ref{fig:dens} for a uniform gas distribution at $10^4$ cm$^{-3}$ irradiated by a sink particle with a Lyman flux of $5\times10^{49}$ s$^{-1}$. In the bottom panel of Figure \ref{fig:dens} we show the convergence with increasing resolution of the shock radius, the green triangles correspond to the particle mass used in the SR simulations. In the top panel we also plot the five previously discussed analytical radii for the ionisation front and shock radius. This test evolves well past the STARBENCH early phase test but ends before stagnation, the point at which an HII region regains pressure balance with its surroundings. This can be seen since Raga-1 does not become flat in Figure \ref{fig:dens} (top panel). We note that our highest resolution run is larger than Raga-II, since they consider the shock to have no thickness, and smaller than Hosokawa-Inutsuka, since they do not consider external pressure.

The accuracy of the numerical approximation to the shock radius is dependent on both the SPH particle mass and the Lyman flux of the source, because of this it is useful to define the resolution relative to both of these quantities. We do this by calculating the number of SPH particles that are equivalent to the gas mass in the Str{\"o}mgren sphere after the R-type expansion phase. We  notice that the accuracy of the shock radius reaches 99\% consistently around the point at which the initial Str{\"o}mgren sphere contains $\approx$ 50 SPH particles. This is not surprising since the nominal resolution in SPH using a cubic spline is 50 particles. In Figure \ref{fig:50sphion} we show the gas densities at which a given Lyman flux will form a Str{\"o}mgren sphere containing mass equal to that of 50 SPH particles for a range of particle masses. 

The Lyman fluxes of sink particles in the SR\_ion simulation vary from $5\times10^{47}$ s$^{-1}$ to $1.3\times10^{51}$ s$^{-1}$. 90\% of sink particles have a flux above $10^{49}$ s$^{-1}$, for which the shock wave radii around ionising sink particles in the SR\_ion run with photoionisation are approximated to an accuracy above 95\%. These estimates are conservative since the $10^4$ cm$^{-3}$ is an upper limit on the typical initial sink surroundings and number densities decrease with radius unlike the test runs. If we consider the total ionising flux, rather than the number of sinks, more than 94\% of the flux contributes to the formation of HII regions which are at least 98\% accurate.  A small number of sink particles in the simulations -- corresponding to a tiny fraction of the overall flux -- are poorly resolved (less than 90\% accuracy). However, we note that many large scale Str{\"o}mgren volume approaches do not resolve individual HII regions at all. These estimates do not consider the cases in which two ionising sinks lie close to one another, in such situations the accuracy would likely be improved.

We also consider the high resolution runs (HR and HR\_ion) and we find that there are no significant statistical differences between these and the fiducial runs. We were only able to run the HR\_ion simulation to 3.5 Myr due to the much higher computational expense. 

%%%%%%%%%%%%%%%%%%%%%%%%%%%%%%%%%%%%%%%%%%%%%%%%%%

% Don't change these lines
\bsp    % typesetting comment
\label{lastpage}
\end{document}